\documentclass[aps,amsmath,amssymb,prb,superscriptaddress,twocolumn]{revtex4}
\usepackage {graphicx}

\newcommand {\be}{\begin {eqnarray}}
\newcommand {\ee}{\end {eqnarray}}

\newcommand {\bR} {{\mathbf R}}
\newcommand {\br} {{\mathbf r}}

\newcommand {\ud} {{\textrm d}}

\begin{document}

\title {Pacifying the Fermi-liquid:  battling the devious fermion signs.}

\author {J. Zaanen}
\email {jan@lorentz.leidenuniv.nl}
\affiliation {Instituut Lorentz voor de theoretische natuurkunde, 
Universiteit Leiden, P.O. Box 9506, NL-2300 RA Leiden, 
The Netherlands}
\author {F. Krueger}
\email {Krueger@lorentz.leidenuniv.nl}
\affiliation {Instituut Lorentz voor de theoretische natuurkunde, 
Universiteit Leiden, P.O. Box 9506, NL-2300 RA Leiden, 
The Netherlands}
\author {J.-H. She}
\email {she@lorentz.leidenuniv.nl}
\affiliation {Instituut Lorentz voor de theoretische natuurkunde, 
Universiteit Leiden, P.O. Box 9506, NL-2300 RA Leiden, 
The Netherlands}
\author {D. Sadri}
\email {sadri@lorentz.leidenuniv.nl}
\affiliation {Instituut Lorentz voor de theoretische natuurkunde, 
Universiteit Leiden, P.O. Box 9506, NL-2300 RA Leiden, 
The Netherlands}
\author {S.I. Mukhin}
\email {sergeimoscow@online.ru}
\affiliation {Moscow Institute for Steel and alloys,Leninski ave. 4, 119049 Moscow, Russia}

\date {\today}
\begin {abstract}
The fermion sign problem is studied in the path integral formalism.
The standard picture of Fermi liquids is first critically analyzed,
pointing out some of its rather peculiar properties.
The insightful work of Ceperley in constructing fermionic path
integrals in terms of constrained world-lines is then reviewed. In this representation,
the minus signs associated with Fermi-Dirac statistics are self
consistently translated into a geometrical constraint structure (the {\em nodal hypersurface})
acting on an effective bosonic dynamics. As an illustrative example we use this formalism to study
$1+1$-dimensional systems, where statistics are irrelevant, and hence
the sign problem can be circumvented. In this low-dimensional example,
the structure of the nodal constraints leads to a lucid picture of the
entropic interaction essential to one-dimensional physics.
Working with the path integral in momentum space, we then show that the
Fermi gas can be understood by analogy to a Mott insulator in a
harmonic trap. Going back to real space, we discuss the topological properties of the
nodal cells, and suggest a new holographic conjecture relating Fermi
liquids in higher dimensions to soft-core bosons in one dimension.
We also discuss some possible connections between mixed Bose/Fermi
systems and supersymmetry.
\end {abstract}

\maketitle

\section {Introduction}

Since the last twenty years or so a serious intellectual crisis has developed in the condensed matter physics enterprise 
dealing with strongly interacting electrons  in solids. An intellectual crisis in science is actually the best
one can hope for -- it should not be confused with an economic- or  sociological crisis. This
field is flourishing right now \cite{science} and there is a general perception that after a slump in the 1990's the field has
reinvented itself.  What is this intellectual crisis about? Substantial progress is made on the experimental side, both with regard
to the discovery of electron systems in solids that behave in very interesting and puzzling ways (high-$T_c$ superconductors\cite{highTc}
and other oxides\cite{mang}, 'heavy fermion' intermetallics\cite{heafe}, organics\cite{organics}, 2DEG's in semiconductors\cite{mitranss}), 
and in the  rapid progress of new instruments 
that make possible to probe deeper and farther in these mysterious  electron worlds (scanning tunneling spectroscopy\cite{sctun}, 
photoemission\cite{ARPES}, neutron- \cite{tranqua} and resonant X-ray scattering\cite{abbamonte}). On the theoretical side there is also much action.
This is energized by the 'quantum field theory'\cite{fradkin} revolution that started in the 1970's in high energy physics,
and is still in the process of unfolding its full potential in the low energy realms, as exemplified by topological
quantum computation\cite{topqucomp}, quantum criticality\cite{sachdev} and so forth.

But the intellectual crisis manifests itself through the fact that the experimental- and theoretical communities are
increasingly drifting apart despite all the pressures to stay together. This is not because the people are bad scientists but
instead it is caused by the dynamics of science itself.  The theorists deduce from their powerful field theory very interesting
suggestions for experiment but these are either impossible to realize in  the laboratory or they are positioned on the fringe
of the experimental main stream. The attitude of the  experimentalists is determined by the expectation that  theorists are just
there to explain why the data do not resemble anything that is found in the books.
 
The cause is obvious. The experimentalists measure systems formed from electrons and electrons are fermions. 
The theorists are playing with the mathematical marvel called quantum field theory.   But the latter works so well because via
the euclidean path integral it boils down to exercises in equilibrium statistical physics. It is about computing probabilistic
partition sums in euclidean space-time following the recipe of Boltzmann and this seems to have no secrets for humanity.  The origin
of the crisis is that this Boltzmannian path integral logic does not work at all when one wants to describe problems characterized
by a finite density of fermionic particles. The culprit is that the path integral is suffering from the fermion sign problem. The Boltzmannian
computation is spoiled by 'negative probabilities' rendering the approach to be mathematically ill defined. In fact, the mathematics 
is as bad as it can be: Troyer and Wiese\cite{troyer} showed recently that the sign problem falls in the mathematical complexity class "NP hard",
and the Clay Mathematics Institute has put one of its 7 one million dollar prizes  on the proof that such problems cannot be solved in 
polynomial time.

Although not always appreciated, the fermion sign problem is quite consequential for the understanding of
the physical world. Understanding  matter revolves around the understanding of  the emergence principles prescribing how
large number of simple constituents (like elementary particles) manage to acquire very different  properties when they
form a wholeness. The path integral is telling us that in the absence of the signs these principles are the same for quantum matter as
they are for classical matter.  But these classical emergence principles are in turn resting on  Bolzmannian statistical physics. When this
fails because of the fermion signs, we can no longer be confident regarding our understanding of emergence. To put it positively,
dealing with fermionic quantum matter there is room for surprises that can be very different from anything we know
from the classical realms that shape our intuition.
In fact, we have only comprehended one such form of fermionic matter: the Fermi-gas, and its "derivative" the Fermi-liquid. The embarrassment
is that we are completely in the dark regarding the nature of other forms of fermionic matter, although we know that they exist because the 
experiments are telling us so.  

The 'quantum weirdness' of the Fermi-gas is obvious: how to understand the Fermi-surface, the Fermi-energy and so forth, just
knowing about classical statistical physics? The interacting Fermi-liquid is a bit more than the Fermi-gas, but focusing on the emergence
principles it is deep inside the same thing. As Landau pointed out, the Fermi-liquid is connected by adiabatic continuation to the
Fermi-gas meaning that the two are qualitatively indistinguishable at the long times and distances where emergence is in full effect. 
The great framework of diagrammatic perturbation theory developed in the 1950's\cite{AGD}  does allow to arrive at quite non trivial statements associated 
with the presence of the interactions but it only works under the condition that the Fermi-liquid is adiabatically connected to the Fermi gas. But conventional
Feynman diagrams are impotent with regard to revealing the nature of 'non Fermi liquids'. To complete the 'fermionic' repertoire of theoretical 
physics,  Bardeen, Cooper and Schrieffer discovered the "Hartree-Fock" mechanism\cite{Anderson}, showing how the Fermi-gas can become unstable towards a 
bosonic state, like the superfluids- and conductors, charge- and spin density wave states and so forth.  Despite fermionic peculiarities (like the gap function),
this is eventually a recipe telling us how the fermi-gas can turn into bosonic matter that is in turn ruled by the Ginzburg-Landau-Wilson classical
emergence rules. In this regard, the other theoretical main streams in correlated electron physics rest on the same bosonization moral:  the one dimensional
electron systems\cite{giamarchi}, as discussed in section VI; the Kondo-type impurity problems being boson problems in disguise\cite{kondo}, which in turn form the fundament for
the popular dynamical mean-field theory\cite{dmft}, and so forth.

Summarizing, given the present repertoire of theoretical physics, all we know to do with fermionic matter is to hope that it is a Fermi gas or bound in bosons.     
But we are facing a zoo of 'non-Fermi-liquid' states of electrons coming out of the experimental laboratories and the theorists are standing empty handed
because the fermion signs render all the fancy theoretical technologies to be useless. The NP hardness of the sign problem tells us that there is no 
mathematically exact solution but how many features 
of the physical world we understand well are actually based on exact mathematics? Nearly all of it is based on an effective description, mathematics that is
tractable while it does describe accurately what nature is doing although it is not derived with exact mathematics from the first principles.  Is there a way to
handle non-Fermi-liquid matter on this phenomenological level?

The remainder of this paper is dedicated to the case that there is reason to be optimistic. This optimism is based on a brilliant discovery some fifteen years 
ago of an alternative path-integral description of the fermion problem by David Ceperley\cite{Cepstatphys,Cepx}. This 'constrained' or 'Ceperley' path integral has a Boltzmannian
structure (i.e., only positive probabilities) but the signs are traded in for another unfamiliar structure: a structure of constraints acting on a 'bosonic' configuration
space that is coding for all the effects of Fermi-Dirac statistics. This is called the reach and it amounts to the requirement that for all imaginary times $\tau$ between
zero and $\hbar \beta$ ($\beta = 1 / (k_B T)$) the worldline configurations should not cross the hypersurface determined by the zero's of the full $N$-particle, imaginary
time density matrix. 
Although the constrained path integral suffers from a self-consistency problem since the exact constrain structure is not known except for the non-interacting Fermi-gas, 
it appears that this path integral is quite powerful for the construction of phenomenological effective theories.  The information carried by the reach lives  'inside'
the functional integral and should therefore be averaged. This implies that only global- and averaged properties of this reach should matter for the physics
in the scaling limit. The reach is in essence a high dimensional geometrical object, closely related to the more familiar 'nodal hypersurface' associated with the sign
changes of ground state wave function.  The theoretical program is  to classify the geometrical and topological properties of the reach in general terms, to find
out how this information is averaged over in the path integral, with the potential to yield eventually a systematic classification of phenomenological theories 
of fermionic matter.

Given that Ceperley derived his path integral already quite some time ago, why is it not famous affair?  These
path integral are not so easy to handle. Although various  interesting results were obtained\cite{CPIresults}, even the attempt to reconstruct the Fermi-liquid in this language
stalled\cite{cepprivat}.  But these efforts were limited to a very small community, with a focus on large scale numerical calculations. The potential of the  Ceperley path integral
to address matters of principle appears to be overlooked in the past. We discovered the Ceperley path integral in an attempt to understand the scale invariant fermionic quantum
critical states as found in the heavy fermion intermetallics.  We started out  on the more primitive level of wave function nodal structure, discovering by accident the much
more powerful Ceperley path integral approach.  We believe that we have delivered proof of principle\cite{krueger} that this language gives penetrating insights in the nature of a prominent 
non-Fermi liquid state: the fermionic quantum critical states realized in the heavy fermion intermetallics.  Since this work is still under review we will not address it in any detail.
However,  to make further progress, we were confronted with the need to better understand the detailed workings of the
Ceperley path integral and we decided to revisit the description of the Fermi gas and the Fermi liquid. The outcomes of this pursuit are summarized in this paper. This paper
contains some new results: the supersymmetric quantum gas as implied by even permutations (\ref{susycycles}),  and especially the closed solution of the Ceperley path integral of the Fermi gas in momentum space (section VII). But there are also many loose ends and 
this paper is in first instance intended as an easy to read tutorial on the Ceperley path integral. We hope  that it will infect others to take up this fresh- and wide open subject, 
where much terrain can be conquered.

This tutorial is organized as follows.
We start out in chapter II with a somewhat unconventional discussion of the Fermi-liquid. To get the problem sharply in focus, we step back from the usual
textbook viewpoint and instead consider the Fermi-liquid from the perspective of the emergence principles governing classical- 
and bosonic matter. In the present context, this perspective has a special relevance: Ceperley's path integral tells that fermionic matter is also subjected to 
the rules of statistical physics and viewed in this light the Fermi-liquid turns into an outrageous, confusing entity. We will make the case that the
Fermi-liquid is {\em holographic} in the same sense of the holographic principle associated with black holes and string theory\cite{AdS}. We claim that regardless the number of 
space dimensions the physics of the Fermi-liquid at low temperature is  in one-to-one relation with the physics of a system of soft core interacting bosons in one dimension.
This might sound absurd but it will turn out to become more reasonable dealing with the real space representation of the Ceperley path integral in the final chapter.      

We continue in chapter III reviewing the only signful fermion path integral that can be solved: the classic Feynman path integral for
the Fermi-gas. This is just a  summary of the beautiful treatment  found in Kleinert's path integral book\cite{Kleinert}. This story appears to be not as widely known as it
should. It shows that the Fermi-gas is quite like the Bose gas, where the hard work is done by worldlines that at low temperature wrap infinite times around the
imaginary time circle. However, the negative probabilities interfere, turning into alternating sums over winding numbers that  eventually take the shape of the 
Fermi-Dirac distribution function.  Chapters IV  is dealing with side lines inspired on section III but
we found both instructive to an extent that they should be included. Chapter IV is actually of some relevance for the Ceperley path integral but it might well have
a broader significance.  We were puzzled by the issue of how to deal with the prescription that only even permutations should be summed over in the Ceperley
path integrals. It is well understood that the braiding properties of worldlines underpin the workings of quantum statistics, referring for instance to the understanding
of anyons and topological quantum computation in two space dimension. What is then the meaning of the even permutations? As we will show in section IV,
it renders the free quantum gas to become supersymmetric!   

In Chapter V the core business starts: we introduce the Ceperley path integral, reviewing it's derivation as well as various other technical issues.  As a first example of
the workings of this path integral we will discuss in section VI the one dimensional Fermi-gas. In a way it is nothing new, but we will make the case that the bosonization step 
becomes particularly transparent in the Ceperley language: the reach becomes  the 'Pauli hypersurface', meaning that the  fermion statistics just  
takes the form of hard core interactions between
the particles. We subsequently highlight a 'maximally' statistical physics view on the one dimensional electron systems that one of us developed some time ago\cite{za1d,zamu1d} but
which is not particularly well known in the community. This emphasizes the aspect that the typical 'fermionic' aspects of physics in one dimensions are actually coding
for the rather intricate effects of entropic interactions, using tricks from soft matter physics to reconstruct the Fermi-gas. 

Chapter VII  is intended to be the highlight of this paper. We present a quite simple solution of the Ceperley path integral for the Fermi-gas:
the Fermi-gas turns out to be  in one-to-one correspondence with a system of cold atoms in an harmonic trap, subjected to a  deep  optical lattice potential such that the
atoms form a perfect Bose Mott-insulator! This can be taken as completely literal, except that this atom trap lives in single particle momentum space instead
of the real space of the atoms.  You might  have already figured out that this is a correct statement, and you might wonder why this is not in the
undergraduate books.  The reason is that one needs the Ceperley path integral to forget once and for all that fermions are incomprehensible.  

Finally in Chapter VIII we turn to the real space description of the Fermi-gas. This was the alley tried by Ceperley and coworkers where they got stuck,
and we have not managed either to get it fully under control.  The perspective on this
issue does change, knowing about the momentum space solution of chapter VII. What is at stake is the structure of duality transformations in the Ceperley
formalism: real- and momentum space dynamics are dual to each other. In order to learn how to address the problems at arbitrary couplings it is important
to understand the duality structure. On the real space side one gets a better view on the richness of the Ceperley path integral. A highlight is the
understanding of the {\em topology} of the reach, based on a conjecture by Ceperley\cite{Cepstatphys} that was recently proven by Mitas\cite{Mitas}. The outcome is that there is no
topological restriction on the windings of the 'Ceperley particle' worldlines as long as these are constructed from triple particle exchanges. We will argue
that the low temperature thermodynamics should be governed by the winding sector and the zero temperature Fermi gas can be viewed as a 
Bose condensate of Ceperley particles. However, the presence of the reach changes radically the winding statistics as compared to the boson case and
it appears that the windings of the Ceperley particles  in {\em any} higher dimension are counted as if they are the windings associated  with soft core bosons 
living in one space dimension -- the literal interpretation of the 'Fermi-liquid holography' introduced in section II.

\section {Know the enemy: the strangeness of the  Fermi-liquid.}

The only exactly solvable many Fermion problem is the non-interacting Fermi-gas. Surely, every student in 
physics knows the canonical solution. Introduce creation and annihilation operators that anti-commute,

\begin{subequations}
\begin{eqnarray}
\{ c^{\dagger}_{\vec{k}}, c_{\vec{k'}} \} & = & \delta_{\vec{k}, \vec{k'}},\\
\{ c^{\dagger}_{\vec{k}}, c^{\dagger}_{\vec{k'}} \}  & = & \{ c_{\vec{k}}, c_{\vec{k'}} \}  =  0,
\label{anticommute}
\end{eqnarray} 
\end{subequations}
and the Hamiltonian is

\begin{equation}
H_0 = \sum_{\vec{k}}  \varepsilon_k c^{\dagger}_{\vec{k}} c_{\vec{k}},
\label{gashamil}
\end{equation}
where $\vec{k}$ is some set of single particle quantum numbers; a representative example is the spinless gas in the 
continuum where $\vec{k}$ represents single particle momentum and $\varepsilon_k =  \hbar^2 k^2 / 2m$. It follows
from standard manipulations that its grand canonical free energy is
\begin{equation}
F_G = - \frac{1}{\beta} \sum_{\vec{k}} \ln \left( 1 + e^{ -\beta ( \varepsilon_{\vec{k}} - \mu ) } \right),
\label{freeen} 
\end{equation}
where $\beta = 1/ ( k_B T )$ and $\mu$ the chemical potential, tending to the Fermi-energy $E_F$ when $T \rightarrow 0$.
The particle number is

\begin{subequations}
\begin{eqnarray}
N & = &  \sum_{\vec{k}} n_{\vec{k}}, \\
n_{\vec{k}} & = & \frac{1} { e^{ \beta ( \varepsilon_{\vec{k}} - \mu ) }  - 1 },
\label{fermidirac}
\end{eqnarray}
\end{subequations}
where  $n_{\vec{k}}$ is recognized as the momentum distribution function. At zero temperature this momentum
distribution function turns into a step function: $n_{\vec{k}} = 1$ for $|\vec{k}| \le k_F$ and zero otherwise where
the Fermi-momentum $k_F = \sqrt{ 2m E_F / \hbar^2}$. The step smears at finite temperature, 
and this is another way of stating the
fact that only at zero temperature one is dealing with a Fermi-surface with a precise locus in single particle momentum space separating 
occupied- and unoccupied states. 

The simplicity of the Fermi-gas is deceptive. This can be highlighted by a less familiar but illuminating argument. As Landau 
guessed correctly\cite{AGD,Anderson}, the Fermi-gas can be adiabatically continued to the interacting Fermi-liquid. The meaning of this 
statement  is that when one considers the system at sufficiently large times and distances and sufficiently
small temperatures('scaling limit')  a state of interacting fermionic matter exists that is physically indistinguishable from the 
Fermi-gas. It is characterized by a sharp Fermi surface and a Fermi energy but now these are formed from a gas of non-interacting
quasiparticles that have still a finite overlap ('pole strength' $Z_{\vec{k}}$) with the bare fermions, because the former are just
perturbatively dressed versions of the latter, differing from each other only on microscopic scales\cite{AGD}. This is the standard lore, 
but let us now consider these matters with a bit more rigor. The term describing the interactions between the bare fermions 
will have the general form,

\begin{equation}
H_1 = \sum_{\vec{k}, \vec{k'} \vec{q}} V ( \vec{k}, \vec{k'}, \vec{q} )    c^{\dagger}_{\vec{k} + \vec{q}} c_{\vec{k}}
 c^{\dagger}_{\vec{k'} - \vec{q}} c_{\vec{k'}}.
\label{interactions}
\end{equation}

It is obvious that single particle momentum does not commute with the interaction term,

\begin{equation}
\left[    c^{\dagger}_{\vec{k}} c_{\vec{k}}, H_1 \right] \neq 0,
\label{nonsinglepartmom}
\end{equation}
henceforth, single particle momentum is in the presence of interactions no longer a quantum number and single particle
momentum space becomes therefore a fuzzy, quantum fluctuating entity. But according to Landau we can still point at
a surface with a sharp locus in this space although this space does not exist in a rigorous manner in the presence of 
interactions! 

In the textbook treatments of the Fermi-liquid this obvious difficulty is worked under the rug. Since the above argument is
rigorous, it has to be the case that the Fermi-surface does not exist when one is dealing with any finite number of particles!
Since we know empirically that the Fermi-liquid exists in the precise sense that interacting Fermi-systems are characterized
by a Fermi-surface that is precisely localized in momentum space in the {\em thermodynamic limit} it has to be that this
system profits from the singular nature of the thermodynamic limit, in analogy with the mechanism of  spontaneously symmetry 
breaking that rules bosonic matter. 

We refer to the peculiarity of bosonic- and classical systems that (quantum) phases of matter acquire a sharp identity only
when they are formed from an infinity of constituents\cite{Anderson,Wezel}. Consider for instance the quantum crystal, breaking spatial translations and
rotations. Surely, one can employ a STM needle to find out that the atoms making up the crystal take definite positions in
space but this is manifestly violating the quantum mechanical requirement that 'true' quantum objects should delocalize over
all of space when it is homogeneous and isotropic. The resolution of this apparent paradox is well known. One should 
add to the Hamiltonian an 'order parameter' potential $V({\bf R})$ where ${\bf R}$ refers to the $dN$ dimensional configuration
space of $N$ atoms in $d$ dimensional space, having little potential valleys at the real space positions of the atoms in the crystal.
It is then a matter of order of limits,

\begin{subequations}
\begin{eqnarray}
\lim_{N \rightarrow \infty} \lim_{ V \rightarrow 0} \langle \sum_i \delta ( \vec{r}_i - \vec{r}^0_i) \rangle & = &  0,\\
\lim_{ V \rightarrow 0} \lim_{N \rightarrow \infty} \langle \sum_i \delta ( \vec{r}_i - \vec{r}^0_i)\rangle & \neq & 0,
\label{singlimit}
\end{eqnarray}
\end{subequations}
where $\vec{r}_i$ and  $\vec{r}^0_i$ are the position operator and the equilibrium position of the $i$-th atom forming
the crystal. Henceforth, the precise positions of the atoms in the solid, violating the demands of quantum mechanical 
invariance, emerge in the thermodynamic limit -- we know that a small number of atoms cannot form a crystal in a rigorous
sense. 

Returning to the Fermi-liquid, the commonality with conventional symmetry breaking is that in both cases non existent 
quantum numbers (position of atoms in a crystal, single particle momentum in the Fermi-liquid) come into existence
via an 'asymptotic' emergence mechanism requiring an infinite number of constituents, at least in principle. But this
is as far the analogy goes. In every other regard, the Fermi-liquid has no dealings with the classical emergence principles,
that also govern bosonic matter.

Although it is unavoidable that the Fermi-liquid needs the thermodynamic limit it is not at all clear what to take for the
order parameter potential $V$. In this regard, the Fermi-liquid is plainly mysterious. The textbook treatises of the
Fermi-liquid, including the quite sophisticated 'existence proofs', share a very perturbative attitude. The best treatments
on the market are the ones based on functional renormalization and the closely related constructive field theory\cite{Salmhofer,Feldman}. 
Their essence is as follows: start out with a Fermi gas and add an infinitesimal interaction, follow the (functional) renormalization
flow from the UV to the IR to find out that all interactions are irrelevant operators. Undoubtedly, the conclusions from these
tedious calculations  that the Fermi-gas is in a renormalization group sense stable against small perturbations are correct.
The problem is that these perturbative treatments lack the mighty general emergence  principles that we worship when dealing 
with classical and bosonic matter. 

To stress this further, let us consider a rather classic problem that seems to be more or less forgotten although it was
quite famous a long time ago: the puzzle of the $^3$He Fermi-liquid\cite{Anderson}. The $^3$He liquid at temperatures in the 
Kelvin range is not yet cohering and it is well understood that it forms a dense van der Waals liquid. Such liquids have
a bad reputation; all motions in such a classical liquid are highly cooperative to an extent that all one can do is to put 
them into a computer and solve the equations of motions by brute force using molecular dynamics. When one cools
this to the millikelvin range, quantum coherence sets in and eventually one finds the impeccable textbook version
of the Fermi-liquid: the macroscopic properties arise from dressed helium atoms that have become completely
transparent to each other, except that they communicate via the Pauli principle, while they are roughly ten times as
heavy as  real $^3$He atoms. When one now measures the liquid structure factor using neutron scattering one
finds out that on microscopic scales this Helium Fermi-liquid is more or less indistinguishable from the classical
van der Waals fluid! Hence, at microscopic scales one is dealing with the same 'crowded disco' dynamics as in the
classical liquid except that now the atoms are kept going by the quantum zero-point motions. On the microscopic
scale there is of course no such thing as a Fermi surface. For sure, the idea of renormalization flow should still apply,
and since one knows what is going on in the UV and IR one can guess the workings of the renormalization flow
in the $^3$He case: one starts out with a messy van der Waals ultraviolet, and when one renormalizes by integrating
out short distance degrees of freedom one meets a 'relevant operator creating the Fermi-surface'. At a time scale
that is supposedly coincident with the inverse renormalized Fermi-energy this relevant operator takes over and drags
the system to the stable Fermi-liquid fixed point. How to construct such a 'Fermi-surface creation operator'? Nobody
seems to have a clue!

Although the microscopic details are quite different, the situation one encounters in interesting electron system like
the ones realized in manganites\cite{mang,mangzx}, heavy fermion intermetallics\cite{heafe} and cuprate superconductors
\cite{highTc} is in gross outlines very similar as in $^3$He. In various guises one finds coherent quasiparticles 
(or variations on the theme, like the Bogoliubons in the cuprates) only at very low energies and low temperatures. 
Undoubtedly the UV in these systems has much more to do with the van der Waals quantum liquid than with a free Fermi-gas. 
Still, the only activity the theorists seem capable off is to declare the UV to be a Fermi-gas that is hit by small interactions. It is 
not because these theorists are incompetent: humanity is facing the proverbial brick wall called the fermion sign problem that
frustrates any attempt to do better.

Arrived at this point we hope that we have convinced the reader that even the 'simple' Fermi-liquid is plainly 
mysterious. This mystery is of course rooted in the fact that the fermion signs disconnect the many fermion
problem from the powerful principles of statistical physics that allow us to fully comprehend the emergence
logic of classical- and bosonic matter. We already stressed that nobody has a clue how to construct a mathematical
definition of a 'Fermi-surface generating renormalization  group operator' and the closely related issue of the
 'Fermi-surface stabilizing order parameter potential'. But there are a couple of other features that are disconnected 
 from anything we know in statistical physics.

The relationship between thermal and quantum fluctuations is plainly weird in Fermi-liquids and Fermi-gasses. 
In sign free, i.e. bosonic or 'Bolzmannion', quantum systems one has a simple rule telling how these fluctuations
relate, which is rooted in the postulates. The thermal path-integral can be taken as basic postulate, being both
applicable to fermionic and sign-free quantum matter\cite{sachdev}. It states that everything takes place in Euclidean space-time,
being spanned by the space dimensions and imaginary time $\tau$. In this formalism, temperature determines
the 'maximum duration of imaginary time': for open spatial boundaries, euclidean space-time has the topology
of a cylinder where imaginary time is the compact direction with a compactification radius $R_{\tau} = \hbar / (k_B T)$.
Henceforth, at zero temperature everything takes place in a $(d+1)$-dimensional space (ignoring complications like
an external heat bath) where $d$ is the number of space dimensions. Addressing general scaling limit issues, like the
existence (or not) of order,  one is at finite temperatures interested in times long compared to $R_{\tau}$.  It follows 
that thermal fluctuations are 'one dimension more important' than quantum fluctuations, at least as long one can get 
away with the well understood role of target space dimensionality in statistical physics. Henceforth, the well known
Mermin-Wagner rules\cite{MerWag} imply that at zero temperature one can have algebraic long range order in $d=1$, while at any 
finite temperature  one needs at least $d=2$; one can truly break a continuous symmetry in $d=2$ at $T=0$, but at any finite
temperature one needs $d=3$, and so forth. This 'space-time geography' applies as well to fermionic problems but
the complication is of course that the connection with statistical physics is shattered! This has a very strange consequence
that can be easily overlooked. We argued already that in some weird fermionic sense, the Fermi-liquid 'breaks symmetry'.   
But from the canonical side we know some answers: from the discontinuity in the momentum distributions we learn that
the Fermi-surface acquires a precise locus in momentum space only at zero temperature (omitting the non-generic $d=1$ 
case that follows the boson rules\cite{giamarchi}). The Fermi-Dirac distribution teaches us in turn that the Fermi-surface
'smears' in momentum space at any finite temperature, regardless the dimensionality of target space. Henceforth, one has
zero temperature order, and finite temperature disorder, regardless dimensionality to the extent that it is even true in
$d = \infty$. This is quite hard to comprehend  when you would only know statistical physics!

We can actually push this further by considering the thermodynamics of the Fermi-liquid in more detail, just forgetting for the moment
how we got there, and insisting that there exist eventually a bosonic/Boltzmannian description. The argument is a no-brainer
but the conclusion is  quite spectacular: the Fermi-liquid demonstrates an extreme form of the {\em holographic principle} that was discovered
in the context of the quantum physics of black holes\cite{AdS}. The precise statement is: {\em the low energy physics of a Fermi-liquid in arbitrary dimensions
is in precise correspondence with an interacting  Boltzmannian system in 1+1 dimensions}. This is surely consistent with the observation
that the Fermi-liquid shows a zero temperature (algebraic) order, while it is disordered at any finite temperature. This is the typical
trait of one dimensional bosonic physics but the Fermi-liquid weirdness is that it is doing this job in all dimensions. Let us make this claim
more precise, by considering the grand canonical free energy of the Sommerfeld  gas in arbitrary space dimension $d>2$. This can be regarded as 
representative for the Fermi-liquid in the scaling limit, i.e. modulo the renormalization of the Fermi-energy and at temperatures 
sufficiently small compared to the Fermi-energy,

\begin{eqnarray}
F_G & = & -\frac{2}{d+2}N E_F\left[1+\frac{\pi^2}{6}\left(\frac{d}{2}+1\right)\left(\frac{k_B T}{E_F}\right)^2+\right.\nonumber\\
& & \left.\mathcal{O}\left(\left(\frac{k_B T}{E_F}\right)^4\right)\right].
\label{freeensom}
\end{eqnarray}

From this free energy follows  the temperature dependence of the specific heat,

\begin{eqnarray}
C_V & = & d N k_B\left[\frac{\pi^2}{6}\frac{k_B T}{E_F}+\mathcal{O}\left(\left(\frac{k_B T}{E_F}\right)^3\right)\right],
\label{spheatsom}
\end{eqnarray}    
and the chemical potential,

\begin{eqnarray}
\mu & = & E_F\left[1-\frac{\pi^2}{12}(d-2)\left(\frac{k_B T}{E_F}\right)^2\right.\nonumber\\
& & +\left.\mathcal{O}\left(\left(\frac{k_B T}{E_F}\right)^4\right) \right].
\label{chpotsom}
\end{eqnarray}    
The Sommerfeld expansion breaks down in $d=2$ where the above thermodynamic functions become non-analytic functions at $T=0$ and cannot 
be expanded in powers of $k_B T/E_F$. For example, for the chemical potential in $d=2$ one obtains $\mu=k_B T\ln[\exp(E_F/k_B T)-1]$. However,
the above expressions strictly hold in $d=2+\epsilon$. 
 
Let us now consider an arbitrary interacting massless bosonic system. In any space dimension $d \ge 1$ such a system cannot avoid (algebraic)
long range order and the thermodynamics is set by the massless Goldstone bosons characterized by a dispersion $\epsilon(k)=c\hbar k$. Assuming 
that the order survives at finite but small temperatures the Free energy becomes,

\begin{equation}
F=-\Gamma_d V k_B T\left(\frac{k_B T}{\hbar c}\right)^d,
\label{freeenbos}
\end{equation}
with $\Gamma_d$ a dimensionless prefactor and it follows that $C_V =d(d+1)\Gamma_d k_B V\left(\frac{k_B T}{\hbar c}\right)^d $, while for the 
chemical potential it is interesting to consider a superfluid where $\mu \sim -T^2$ for $1+1$D Boson systems.

As we learned in freshmen courses, the temperature dependence of thermodynamic quantities of Boltzmannions strongly depends on temperature,
like the Debye specific heat $C_V \sim T^d$, reflecting that for increasing dimensionality more collective degrees of freedom become available
with the effect that entropy increases more rapidly for increasing temperature. On the other hand, {\em the number of degrees of freedom counted 
by the increase of entropy for increasing temperature of  the  Fermi-liquid is entirely independent of dimensionality}, 
being actually coincident with the number of degrees of freedom of a 1+1D bosonic system. We learned to comprehend this on the signfull side by
arguing that the microscopic degrees of freedom are locked up in the Fermi sea, while for rising temperature only degrees of freedom are released
in a thin shell $\sim k_B T / E_F$ around the Fermi-surface. This is of course a fine explanation but to make it work we need the fermion signs. But
we have now learned that a Boltzmannian description of the Fermi-liquid exists, in the form of the Ceperley path integral. Although the constrained
structure is quite non-trivial, it cannot cause miracles and because it is a Boltzmannian machine it has to give in eventually to the iron 'Mermin-Wagner'
order parameter rules. Henceforth, it has to be that in the Ceperley formalism we are dealing with an effective 1+1 dimensional order parameter theory.

The last 'anomaly' of the Fermi-liquid appears again as rather innocent when one has just worked oneself through a
fermiology textbook. However, giving this a further thought,  it is actually the most remarkable and most mysterious 
feature of the Fermi-liquid. Without exaggeration, one can call it a 'UV-IR connection', indicating 
the rather unreasonable way in which microscopic information is remembered in the scaling limit. It refers to the well known
fermiology fact that by measuring magneto-oscillations in the electrical transport (Haas van Alphen-, and Shubnikov 
de Haas effects) one can determine directly the average distance between the microscopic fermions by executing 
measurements on a macroscopic scale\cite{AsMer}. This is as a rule fundamentally impossible in strongly interacting classical-
and sign free quantum matter. Surely, this is possible in a weakly interacting and dilute classical gas, as used with 
great effect by van der Waals in the 19-th century to proof the existence of molecules. But the trick does not work in
dense, strongly interacting classical fluids: from the hydrodynamics of water one cannot extract any data regarding
the properties of water molecules. Surely, the weakly interacting Fermi-gas is similar to the van der Waals gas but
a more relevant example is the  strongly interacting $^3$He, or either the heavy fermion Fermi-liquid. At microscopic
scales it is of course trivial to measure the inter-particle distances and the liquid structure factor of $^3$He will directly
reveal that the helium atoms are apart by $4$ angstroms or so. But we already convinced the reader that there is
no such thing as a Fermi surface on these scales. Descending to the scaling limit,  a Fermi-surface emerges and it encloses
a volume that is protected by the famous Luttinger theorem\cite{lutt,oshi}: {\em it has to enclose the same volume as the non-interacting
Fermi gas at the same density!} Using macroscopic magnetic fields, macroscopic samples and macroscopic distances 
between the electrical  contacts one can now measure via de Haas van Alphen effect, etcetera, what $k_F$ is and the Fermi
momentum is just the inverse of the inter-particle distance modulo factors of $2 \pi$. This is strictly unreasonable. We
repeat, on microscopic scales the system has knowledge about the inter-particle distance but there is no Fermi-surface;
the Fermi surface emerges on a scale that is supposedly in some heavy fermion systems a factor 100 or even 1000
larger than the microscopic scale. But this emerging Fermi-surface still gets its information from somewhere, so that
it knows to fix its volume satisfying Luttinger's rule!  In a later section we hope to shed some light on the 'mysteries' 
addressed in this section using Ceperley's path integral but we are still completely in the dark regarding this particular
issue. It might well be that there are even much deeper meanings involved; we believe that it has dealings with the
famous anomalies in quantum field theories\cite{wein}. These are tied to Dirac fermions and the bottom line is that these process
in rather mysterious ways ultraviolet (Planck scale) information to the infrared, with the effect that a gauge symmetry
that is manifest on the classical level is destroyed by this 'quantum effect'.   

To summarize, in this section we have discussed the features of the Fermi-liquid that appear to be utterly mysterious 
to a physicist believing that any true understanding of physics has to rest on Boltzmannian principle:\\
(i) What is the order parameter and order parameter potential of the zero temperature Fermi-liquid?\\
(ii) How to construct a 'Fermi-surface creation operator', which is supposed to be the relevant operator associated with
the IR stability in the renormalization group flow?\\
(iii) Why is there 'Fermi-liquid order' at zero temperature in any $d \ge 2$, while it gets destroyed by any finite temperature
regardless dimensions? More precisely, why is the Fermi-liquid holographic and what are the degrees of freedom of the 'holographic 
screen' populated by the effective bosons? \\
(iv) Why is it possible to retrieve microscopic information via the Luttinger sum rule by performing macroscopic magneto-transport
measurements, even in the asymptotically strongly interacting Fermi-liquid?

The bottom line of this paper will be, although we know for sure that there are 'Bolzmannian' answers to these questions, and 
although we know quite well where to look for them, we have no conclusive answers in the offering right now. But the remainder
will make clear why these questions are so interesting.

\section{The sign-full worldline path integral.}

There is just one sign-full path integral problem that can be completely solved: the non-interacting Fermi gas in
worldline representation, in any dimension. It is the usual business, when one can solve a problem exactly in
one representation (i.e. canonical) it can also be solved in any other representation. In fact, the Fermi-gas path
integral is a textbook problem, although we are aware of only one textbook where it is worked out in detail: Kleinert's
Path integral book\cite{Kleinert}.   Let us first summarize the 'mechanics' of this path-integral, referring the reader to Kleinert's book
for the details, to subsequently use this solvable case as an example to highlight the rather awkward and counterintuitive
workings of the 'negative probabilities'. All along it is interesting to compare it with the free boson path integral which works
the same way except that it corresponds with a well behaved Bolzmannian problem.

Consider the partition function for Bosons or Fermions; this can be written as an integral over configuration
space $\bR = (\br_1,\ldots, \br_N) \in \mathbb{R}^{Nd} $ of the diagonal density matrix evaluated at an imaginary $\hbar\beta$,
\begin{equation}
\label{partition-function-1}
\mathcal{Z} =  \textrm{Tr} e^{- \beta H} = \int \ud\bR \rho (\bR,\bR;\beta).  
\end{equation}

The path integral formulation of the partition function rests on a formal analogy between the
quantum mechanical time evolution operator in real time
$e^{- i \hat{H} t / \hbar}$ and the finite temperature quantum statistical density operator
$\hat{\rho} = e^{- \beta \hat{H}}$, where the inverse temperature $\beta = 1/k_B T$ has to be identified 
with the imaginary time $it/\hbar$.  The partition function defined as the trace of this operator and expression \eqref{partition-function-1} 
simply evaluates this trace in position space. More formally this can viewed as a Wick rotation of the quantum mechanical path integral, and
requires a proper analytic continuation to complex times.
This rotation tells us that the path integral defining the partition function lives in
$D$-dimensional Euclidean space, with $D=d+1$ and $d$ the spatial dimension of the equilibrium system.
This analogy tells us that to study the equilibrium statistical mechanics of a quantum system
in in $d$ space dimensions, we can study the quantum system in a Euclidean space of dimension
$d+1$, where the extra dimension is now identified as a "thermal" circle of extent $\beta$.
At finite temperature this circle is compact and world-lines of particles in the many-body
path integral \eqref{partition-function-1}  then wrap around the circle, with appropriate
boundary conditions for bosons or fermions.
The discrete Matsubara frequencies that arise from Fourier transforming modes on this circle
carry the idea of Kaluza-Klein compactification to statistical mechanics.
We will come back to a careful consideration of the evaluation of the partition function
\eqref{partition-function-1} in section IV when we discuss the connection between winding
vs. cycle decomposition in preparation for some observations regarding supersymmetry.

For distinguishable particles interacting via a potential $V$ the density matrix can be written in a worldline path integral form as,

\begin{subequations}
\begin{eqnarray}
\rho_D (\bR,\bR';\beta) & = &  \int_{\bR\to\bR'} \mathcal{D}\bR  \exp(-\mathcal{S}[\bR]/\hbar),\\
\mathcal{S}[\bR] & = & \int_0^{\hbar \beta} d \tau \left( \frac{m}{2} \dot{\bR}^2 (\tau) + V (\bR(\tau) ) \right),\quad  
\end{eqnarray}
\label{pathdist} 
\end{subequations}
but for indistinguishable bosons or fermions one has also to sum over all $N!$ permutations ${\cal P}$ of the particle coordinates,
\begin{equation}
\rho_{B/F} (\bR,\bR;\beta ) = \frac{1}{N!} \sum_{\cal P} (\pm 1)^p \rho_D (\bR, {\cal P}\bR; \beta ), 
\label{pathstat} 
\end{equation}
where $p$ is he parity of the permutation. For the bosons one gets away with the positive sign, but for fermions the contribution of a permutation with uneven
parity to the partition sum is a 'negative probability', as required by the anti-symmetry of the fermionic density matrix. This is the origin of the sign problem. 
  
\begin{figure}
\begin{centering}
\includegraphics[width=0.7\linewidth]{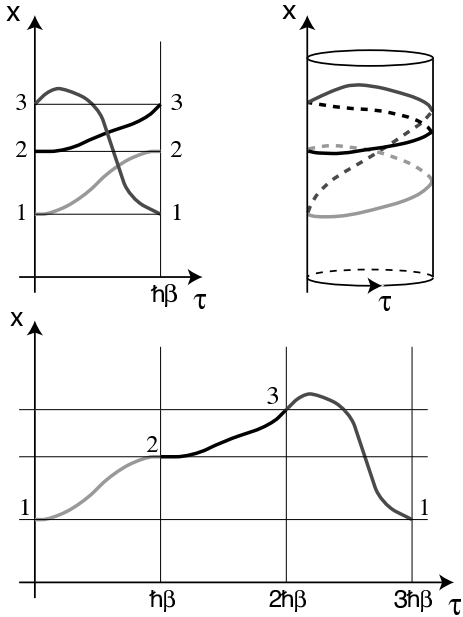}
\end{centering}
\caption{Worldline configuration corresponding to a cyclic exchange of three particles, $1\to2$, $2\to3$, and $3\to1$, or in short notation $(123)$ (upper left).
On a cylinder (upper right), the worldlines form a closed loop winding $w=3$ times around the cylinder.
 In the extended zone scheme (bottom), the exchange process of three particles can be identified with a worldline of a single particle at an effective inverse temperature
 $3\beta$.}
 \label{fig.winding}
\end{figure}

The partition sum describes worldlines that 'lasso' the circle in the time direction. Every permutation in the sum is composed out of so called permutation
cycles. For instance, consider three particles. One particular contribution is given by a cyclic exchange of the three particles corresponding with a single worldline 
that winds three times around the time direction with winding number $w=3$ (see Fig. \ref{fig.winding}), a next class
of contributions correspond with a 'one cycle' with $w=1$ and a two-cycle with $w=2$ (one particle returns to itself while the other two particles are exchanged), and 
finally one can have three one cycles (all particles return to their initial positions). For bosons this is just equivalent
to a problem of interacting ring polymers and this can be solved to any required accuracy using quantum Monte-Carlo -- see e.g. the impressive work by Ceperley on
$^4$He,\cite{ceperhelium,ceperhelium1} making the case that this strongly interacting boson problem has no secrets left. But for fermions one can only handle the non-interacting limit, 
because of the fermion signs.

For the non-interacting Bose and Fermi gas the evaluation of the path integral reduces to a combinatorial exercise. Let us first  illustrate these matters for the example
of $N=3$ particles. It is straightforward to demonstrate, that the identity permutation gives a contribution $Z_0(\beta)^3$ to the partition function (here $Z_0(\beta)$
denotes the partition function of a single particle), whereas an exchange of all three particles contribute as $Z_0(3\beta)$. The meaning is simple: in the absence of interactions
the 3-cycle can be identified with a single particle worldline returning to its initial position at an effective inverse temperature $3\beta$ (see Fig. \ref{fig.winding}). Further on,
a permutation consisting of a $w=1$ and a $w=2$ cycle contributes with $Z_0(\beta)Z_0(2\beta)$. To write down the canonical partition function for $N=3$ non-interacting
bosons or fermions we only have to know the combinatorial factors (e.g. there are 3 permutations made out of a  $w=1$ and a $w=2$ cycle) and the parity of the permutation to 
obtain

\begin{eqnarray}
Z_{B/F}^{(N=3)}(\beta) &  = &   \frac{1}{3!}[Z_0(\beta)^3\pm 3 Z_0(\beta)Z_0(2\beta)\nonumber\\
& & +2 Z_0(3\beta)].
\end{eqnarray}
This result can easily be generalized to $N$ particles. We denote the number of 1-cycles, 2-cycles, 3-cycles, $\ldots N$-cycles the permutation is build of with $C_1$, $C_2$, $C_3$,$\ldots$,
 $C_N$ and denote the combinatorial factors counting the numbers of permutations with the same cycle decomposition $C_1,\ldots C_N$ with $M(C_1,\ldots C_N)$. For $N$ particles
we have to respect the overall constraint $N=\sum_w C_w$ and obtain

\begin{eqnarray}
Z_{B/F}^{(N)}(\beta) &  = &  \frac{1}{N!}\sum_{C_1,\ldots C_N}^{N=\sum_w C_w} M(C_1,\ldots C_N)(\pm 1)^{\sum_w (w-1) C_w}\nonumber\\
& & \times\prod_{w=1}^N\left[Z_0(w\beta)\right]^{C_w}.
\label{ZN}
\end{eqnarray}

Although the combinatorial factors can be written down in closed form,

\begin{equation}
M(C_1,\ldots C_N)=\frac{N!}{\prod_w C_w! w^{C_w}},
\label{comb-factors}
\end{equation}
the canonical partition function (\ref{ZN}) is very clumsy to work with because of the constraint acting on the sum over cycle decompositions. However, it is possible to
derive a recursion relation for the canonical partition function or examine it in terms of so called loop decompositions. For details we refer the reader to 
the appendix. The constraint problem can be circumvented by going to the grand-canonical ensemble. After simple algebraic manipulations we arrive at the grand-canonical partition function

\begin{eqnarray}
Z_{G}(\beta,\mu) &  = &  \sum_{N=0}^{\infty}Z_{B/F}^{(N)}(\beta)e^{\beta\mu N}\nonumber\\
& = & \exp\left( \sum_{w=1}^\infty (\pm 1)^{w-1}\frac{Z_0(w\beta)}{w}e^{w\beta\mu}\right),                   
\label{ZG}
\end{eqnarray}
corresponding to a grand-canonical free energy

\begin{eqnarray}
F_G (\beta) & = & - \frac{1}{\beta}\ln Z_{G}(\beta,\mu)\nonumber\\
 & & = - \frac{1}{\beta} \sum_{w=1}^{\infty} (\pm 1)^{w-1} \frac{ Z_0 (w \beta)} {w} e^{\beta w\mu},
\label{signfreeen}
\end{eqnarray}
with the $\pm$ inside the sum referring to bosons ($+$) and fermions ($-$), respectively. This is a quite elegant result: in the grand-canonical ensemble 
one can just sum over worldlines that wind $w$ times around the time axis; the cycle combinatorics
just adds a factor $1/w$ while $Z_0 (w \beta) \exp{ ( \beta w \mu ) }$ refers to the return probability of a single worldline of overall length $w \beta$. 
In the case of zero external potential we can further simplify 

\begin{eqnarray}
Z_0 (w \beta) & = &  \frac{V^d}{\sqrt{ 2\pi \hbar^2 w \beta / M}^d} \nonumber \\
                         & = & Z_0 (\beta) \frac{1}{w^{d/2}},
\label{zodef}
\end{eqnarray}
to obtain for the free energy and average particle number $N_G$, respectively,

\begin{subequations}
\begin{eqnarray}
F_G & = & - \frac{Z_0(\beta)}{\beta} \sum_{w=1}^{\infty} (\pm 1)^{w-1} \frac{e^{\beta w\mu}}{w^{d/2+1}}, \\
N_G & = & -\frac{\partial F_G}{\partial \mu}=Z_0 (\beta) \sum_{w=1}^{\infty} (\pm 1)^{w-1} \frac{e^{\beta w\mu}}{w^{d/2}}.
\end{eqnarray}
\label{freefreeen}
\end{subequations}

To establish contact with the textbook results for the Bose- and Fermi-gas one just needs that the sums over windings can be written in
an integral representation as,

\begin{equation}
\sum_{w=1}^{\infty} (\pm 1)^{w-1} \frac{e^{\beta w\mu}}{w^{\nu}} = \frac{1}{\Gamma(\nu)} \int_0^{\infty} d \varepsilon 
\frac{\varepsilon^{\nu-1}} {e^{\beta (\varepsilon - \mu)} \mp 1},
\label{fdwinding}
\end{equation}
and one recognizes the usual expressions involving an integral of the density of states ($N(\varepsilon) \sim  \varepsilon^{d/2}$ in $d$ space dimensions) 
weighted by Bose-Einstein or Fermi-Dirac factors.

The bottom line is that at least for bosons Eq. (\ref{freefreeen}) has the structure of a Bolzmannian partition sum and one can rest on the powerful conceptual
machinery of statistical physics. Surely, the canonical route is shorter but it has the 'transfer-matrix attitude': very powerful when it solves the problem exactly
but it is 'overly algebraic', not adding much to the 'no-nonsense' conceptual structure, characteristic for statistical physics.

For example, Einstein just used  Eq.'s (\ref{freefreeen},\ref{fdwinding}) to deduce that for $d \ge 3$ a true phase transition occurs in the Bose gas from a
high temperature classical gas to the Bose-Einstein condensate. Given that there are no interactions this is peculiar and just knowing about the canonical
side one can only take Bose-Einstein condensation as a mysterious quantum phenomenon that drops out from the algebra. But knowing the path-integral
side one cannot afford mysteries: it is just an equilibrium ring polymer problem, and plastic cannot have secrets! Indeed, in the winding representation one
meets a meat-and-potato thermodynamic singularity. At the transition $\mu \rightarrow 0$ and one directly infers from  Eq. (\ref{freefreeen}) that very long
worldlines corresponding with winding numbers $w \sim N$ are no longer penalized, while there are many more long winding- than short winding  contributions
in the sum. It is straightforward to show that in the thermodynamic limit worldlines with $w$ between $\sqrt{N}$ and $N$ have a vanishing weight above the 
BEC temperature, while these infinite long lines dominate the partition sum in the condensate\cite{bund+99}. 

A related issue is the well known fact that the non-interacting Bose-Einstein condensate and the superfluid that occurs in the presence of finite repulsions are
adiabatically connected: when one switches on interactions the free condensate just turns smoothly into the superfluid and there is no sign of a phase transition.
This can be seen easily from the canonical Bogoliubov theory\cite{bogo,fetwal}. Again, although the algebra is fine matters are a bit mysterious. The superfluid breaks 
spontaneous $U(1)$ symmetry, thereby carrying rigidity as
examplified by the fact that it carries a Goldstone sound  mode  while it expels vorticity.  The free condensate is a non-rigid state, that does not break symmetry
manifestly, so why are they adiabatically connected? The answer is obvious in the path-integral representation\cite{ceperhelium,ceperhelium1}: although interactions will hinder the free
meandering of the polymers, a lot of this hindrance is required to make it impossible for worldlines to become infinitely long below some temperature.  The fraction
of infinitely long worldlines is just the condensate fraction and even in the very strongly coupled $^4$He superfluid these still make up for roughly $30 \%$ of all
worldlines! The only way one can get rid of the infinite windings in the interacting system is to turn it into a static array of one cycles - the $^4$He crystal.   This simple
argument underlies the widespread believe that 'simple' bosons can only form superfluids or crystals, while for instance a non-superfluid 'Bose metal'  cannot possibly
exist\cite{phillips}. How to avoid the windings when the worldlines can meander over infinite distances?

This preceding paragraph illustrates the reasons to worship path-integrals when one has learned the language. Are they helpful dealing with fermions?
Let us attempt to address the 'mysteries' (i) and (iii) of Section II: the Fermi-gas represents a form of order (the locus of the Fermi surface, the jump in $n_k$) but 
in all dimensions $d \ge 2$ this 'order' disappears at any finite temperature. Thinking in a statistical physics language it appears at first sight  that the only 
source of this thermodynamic
singularity can reside in the 'infinite windings'. Let us first consider the zero temperature case,where the Fermi gas is described by the wave function 

\begin{equation}
\Psi=\frac{1}{\sqrt {N!}}\det \left( \begin{array}{ccc}
e^{ik_1x_1} & {\cdots} & e^{ik_Nx_1} \\
\vdots & & \vdots \\
e^{ik_1x_N} &  {\cdots} & e^{ik_Nx_N} \end{array} \right),
\end{equation} 

Here the momenta $k_1,\cdots,k_N$ fill the Fermi sea. The slater determinant is a signful summation over $N!$ different permutation patterns, each characterized by 
a certain cycle decomposition as explained previously,  

\begin{eqnarray}
 \Psi&=&\frac{1}{\sqrt{N!}}\sum_\mathcal{P} (-1)^p e^{ik_1x_{p(1)}}  {\cdots}  e^{ik_Nx_{p(N)}}\nonumber\\
 &=&\frac{1}{\sqrt{N!}}\sum_\mathcal{P}\psi_\mathcal{P}.
\end{eqnarray}

Now we want to ask the question, what is the probability for large cycles to occur? This probability can be written as an incoherent sum 

\begin{equation}
P_L=\frac{1}{N!}\sum_{\rm large-cycles} \vert \psi_\mathcal{P}\vert^2,
\end{equation}
which is insensitive to the fermion sign, thus leading to the same result as for bosons.
$P_L$ can be computed by examining the cycle structure of the random permutations of $N$ objects. This is already done by the mathematicians Golomb and Gaal\cite{Gol}. According to their result, 
the probability that the greatest cycle length is $k$ satisfies the recursion relation 

\begin{eqnarray}
P(k,N) & = & \sum_{j=1}^{[N/k]}\frac{1}{j!k^j}\frac{N!}{(N-kj)!}\nonumber\\
& & \times\sum_{t=1}^{{\rm min(k-1,N-kj)}}P(t,N-kj),
\end{eqnarray}
where $[a]$ represents the largest integer not greater than $a$. For $N/2<k\leq N$, the probability has the simple form $P(k,N)=1/k$. For large $N$, the probability that the length 
of greatest cycle lies between $N/2$ and $N$ is

\begin{equation}
\sum_{k=N/2}^{N}P(k,N)\approx \ln 2\approx 0.693.
\end{equation}

For $N/3\leq k<N/2$, the total probability is $\sum_{k=N/3}^{N/2}P(k,N)\approx 0.258$, while for $N/4\leq k<N/3$, $\sum_{k=N/4}^{N/3}P(k,N)\approx 0.044$. Thus there is $99.5$ per cent 
possibility that the greatest cycle has length larger than $N/4$. Our conjecture is that in the large $N$ limit, the probability to have infinite winding is unity 

\begin{equation}
\mathop {\lim }\limits_{N \to \infty }\sum_{k=N_c}^{N} P(k,N)=1,
\label{lc}
\end{equation}
where the lower bound $N_c\to \infty$ as $N\to\infty$, and $\mathop {\lim }\limits_{N \to \infty }N_c/N\to 0$. For example, $N_c$ can be chosen as $\sqrt N$ and one indeed finds that Eq. \eqref{lc}
is satisfied.

So we have rigorously proved that the zero temperature Fermi gas is characterized by infinite windings. As for bosons, does this mean that at any finite temperature 
a sudden change has occurred to short windings? One can easily deduce the answer from Eq. (\ref{freefreeen}). At a temperature of order of the Fermi temperature $T_F$
the chemical potential of the Fermi-gas switches from negative to positive and this means that in any term of the winding sum a worldline which has a winding number
that is larger by one unit than another is more important for the free energy by a factor $e^{\beta \mu}$, a very large factor indeed for $\mu \simeq E_F$ and $\beta
\rightarrow \infty$! Henceforth, when the effects of Fermi-Dirac statistics become noticeable at temperatures $T \le T_F$ the sum is by default dominated by infinite long windings
that however cancel each other nearly completely because of the alternating sign in Eq.  (\ref{freefreeen})!
    
The resolution of this 'paradox' of course lies in the fact that when one is dealing with an alternating sum which is not absolutely convergent every individual term in the sum is meaningless, 
while one has to consider the sum as a whole. To cite a well known example,

\begin{eqnarray}
h & = & 1 - 2 + 4 - 8 + 16 - \cdots \nonumber \\
   & = & 1 - 2h, \nonumber                      
\end{eqnarray}
and it follows that $h = 1/3$. This is of course the same thing as a Fermi-Dirac factor, and after performing the transformation to the integral representation, 
 Eq. (\ref{fdwinding}), one just discovers that the sum over winding boils down to the usual result that the Fermi-Dirac distributions turns  into a step function only when
 $\beta \rightarrow \infty$. The fermion signs translate in the Fermi-Dirac alternating winding sums in the case of the free Fermi gas and that is the
 end of the story. One can only handle these sums when the problem is exactly solvable, and the signs have the net effect  of destroying the connection
 with the powerful conceptual structure of statistical physics.    

The Fermi-gas is to an extent  misleading: it is actually the only example of a fermion system that is completely solved. No closed form, exact solution exists
for finite interactions. Surely, there is no doubt that the Fermi-liquid exists but there is no general mathematical proof stating that it has to exist given
specific microscopic circumstances. The claims in this direction are all based on special, or even pathological limits, like the weak coupling and low
density limits. The Fermi-liquid is at the same time a monument of scaling limit phenomenology\cite{Salmhofer,Feldman}: considering long times and large distances, physics
simplifies and emergence principles take over that can be caught in simple but powerful mathematical structures. To find such principles for fermionic
matter is the outlook of this paper.

To conclude this exposition of the sign problem, the best way to highlight its severity is by employing the rigorous language of mathematical complexity
theory. One can classify mathematical problems according to the criterium of how the computation time of some hypothetical computer will scale with
the number of degrees of freedom $N$. When this time is polynomial in $N$ the problem is considered as solvable in principle. This is different for
problems in the 'nondeterministic polynomial' (NP) class, that have a time that grows exponentially with $N$ on a classical, deterministic computer,
and a special subclass of such problems are called 'NP hard' if any problem in NP can be mapped onto it with polynomial complexity. The bottom line
is that when one can solve one particular NP-hard problem, one has solved all NP problems. To illustrate the gravity of this affair: the traveling salesman and
the spin glass problem are NP hard. Recently, Troyer and Wiese\cite{troyer} proved that the fermion sign problem is NP hard. Hence, when you manage to demonstrate
a polynomial time solution for the sign problem you can cash a million dollars at the Clay Mathematics Institute!

This proof is actually remarkably simple. The classical Ising spin glass in three dimensions,
\begin{equation}
H = - \sum_{<j,k>} J_{jk} \sigma^z_j \sigma^z_k
\label{spglass}
\end{equation}
where the spins $\sigma_j$ take the values $\pm 1$ while the couplings $J$ are randomly chosen from $0, \pm 1$, is the complexity class NP hard. By
a trivial rotation of the quantization axis one can write this as well as a signful quantum problem,
\begin{equation}
H = - \sum_{<j,k>} J_{jk} \sigma^x_j \sigma^x_k
\end{equation}
the Hamiltonian has only off diagonal matrix elements in a basis spanned by eigenstates of $\sigma^z$.  When the exchange couplings $J$ would be all
positive definite all matrix elements would be negative and the ground state wave function would be nodeless and therefore bosonic. This problem is easy
to solve in polynomial time. However, when the signs of the $J$'s can be positive and negative the signs appear, while  at the same time the frustrations
switch on causing the NP-hardness of the equivalent spin-glass problem: Q.E.D.   

\section{Supersymmetry}
\label{susycycles}

In this section we will consider the meaning of supersymmetry in the world-line path integral formalism.
It is a new symmetry introduced to related the physics of bosons (which mediate forces) and
fermions (the constituents of matter) \cite{Wess:1992cp}, and leads to many beautiful mathematical properties
\cite{Terning:2006bq,Weinberg:2000cr}.
Supersymmetry is an idea that has a long and illustrious history in the high-energy physics
community and has made appearances even within condensed matter.
It is presently the most promising idea for new physics beyond the standard model and there is much
hope that the first glimpses of it will be gleaned at the Large Hadron Collider in the next few years.
This hope is predicated on the ability of supersymmetry to solve one of the principal open issues
facing high-energy physics (aside from gravity), that of the hierarchy problem associated with
the mass of the Higgs boson, which is believed to drive the electroweak symmetry-breaking phase
transition. The solution derives from a new class of non-renormalization theorems controlling
certain quantum corrections in manifestly supersymmetric systems.
In the context of gravitational physics, supersymmetry has been instrumental in formulating a consistent
theory of quantum gravity within the framework of string theory, and has led to the discovery of deep
and profound dualities relating seemingly incongruous theories.

It is standard practice when teaching quantum field theory to high-energy physicists these days
to focus on the second quantization of theories at zero temperature and chemical potential, this
being both a convenient limit and a good starting point for those interested in the kind of
collider physics that is the experimental underpinning of high-energy particle physics today.
There are, though, a new generation of experiments, for example the Relativistic Heavy Ion Collider
(RHIC) \cite{Ollitrault:2006va,Yagi:2005yb}, which mean to probe our understanding of Quantum Chromodynamics (QCD, the theory of quarks) at ever higher temperatures and densities, with an eye to understanding the phase transition in the early universe that is conjectured to have transformed a plasma of quarks and gluons into the Hadrons  (protons, neutrons and so on) that dominate the low energy world we see.

Yet, given the prominent role of phase transitions and symmetry-breaking in our understanding of
the standard model and extensions thereof, together with the postulated role of supersymmetry,
a natural question to ponder is the meaning of supersymmetry at finite temperature and density.
It is often stated that supersymmetry is broken at any finite non-zero temperature, though the
precise meaning of this statement and the consequent implications are not immediately obvious.
This belief relies on how supersymmetry relates bosonic and fermionic degrees of freedom and
the fact that at finite temperatures they follow different statistical distributions.
Studies of this question \cite{Moshe:2002ra,Hiller:2006gg,Lucchesi:1998eb,Kratzert:2003cr,Kye:1992ih}
leave open some unresolved questions, in particular the
influence of non-renormalization theorems when zero-temperature supersymmetric systems are
raised to a finite temperature.

We will not address these issues in full generality here, but will point out an interesting
observation (hitherto unknown) about free supersymmetric systems when formulated in the
language of first-quantized world-line path integrals, and suggest ways to press into the
regime of interacting systems.
To set the stage for this discussion, we will first take a detour to study the partition function
of a free system in the world-line language, making explicit the sum over windings. We then
demonstrate that the combinatoric sums can be reorganized into sums over numbers of cycles.
This will be our goal for section \ref{WvsC}.
With this tool kit at our disposal, in section \ref{susycycles}
we then show how supersymmetry can be understood as a restriction
on the types of cycles we must sum over when constructing partition functions.

%\subsection{Odd Windings Lead to Supersymmetry}
We will now present some thoughts on the nature of supersymmetry in the language of world-line path integrals.
Though we only consider the case of free particles explicitly, we feel that this way of looking at supersymmetry might
suggest new insights into the underlying meaning of supersymmetry in a way that allows one to move to study its properties
and implications at finite temperature and chemical potential.

We start by considering the physics of a gas of bosons intermingled with a gas of fermions.
The grand canonical free energy for a gas of free bosons, written as a sum over windings, is
\begin{equation}
  F_G^b \: = \: - \: \frac{1}{\beta} \: Z_0 (\beta)
  \sum_{w=1}^\infty \: \frac{e^{w \beta \mu}}{w^{\frac{d}{2}+1}},
\end{equation}
with $Z_0(\beta)$ the partition function of a single particle at inverse temperature $\beta$, and $D$ is the spatial
dimensionality of the system. For free fermions
\begin{equation}
  F_G^f \: = \: - \: \frac{1}{\beta} \: Z_0 (\beta)
  \sum_{w=1}^\infty \: (-1)^{(w-1)} \: \frac{e^{w \beta \mu^\prime}}{w^{\frac{d}{2}+1}},
\end{equation}
where the chemical potentials are in general unrelated, and consistency requiring that the chemical potential of free
bosons be negative semi-definite. Since no such constraint exists for free fermions, we are free to choose the chemical
potential of the fermions such that $\mu^\prime = \mu \le 0$.
The free energy of the full system then becomes
\begin{equation}
  F_G \: = \: - \: \frac{2}{\beta} \: Z_0 (\beta)
  \sum_{w \ odd} \: \frac{e^{w \beta \mu}}{w^{\frac{d}{2}+1}}
\end{equation}
Here we notice an intriguing property of a gas of free bosons and fermions with equal chemical potential: it is equivalent
(due to cancellations) to a system composed of a new type of particle, with the same chemical potential, but the strange
property that it can only wind an odd number of times around the thermal circle.
Going back to the original picture in terms of a gas of free fermions and bosons, we note that, in the zero temperature limit,
it is now easy to check the expected value of the internal energy vanishes.
The vanishing of the energy of the system is precisely the order parameter for unbroken supersymmetry.
Thus a system of free particles which are only allowed to wind an odd number of times is supersymmetric!
This is true even for finite chemical potential.
The vanishing of the energy at zero temperature is of course also a feature of the normal Bose-Einstein condensate.
This new class of particle which only experiences odd windings also undergoes a condensation in dimensions greater than two, but with a critical temperature that
is shifted higher relative to the normal boson case. At temperatures below this critical value, the chemical potential of the system vanishes (for $d>2$),
which implies that the Fermi surface for the original fermions also degenerates and the fermion occupation vanishes since the fermions
can not condense in the zero-momentum state.
As we cross above the critical temperature the chemical potential becomes finite, a Fermi surface appears,
but simultaneously we have the situation that the occupation of the zero momentum state becomes non-macroscopic.
So magically the appearance of a finite density of fermions is associated with the change from a macroscopically occupied zero-momentum state
to non-macroscopic occupation. The fermions kick out the bosons!

We have arrived at this conclusion without any mention
of an underlying algebraic system describing fermionic symmetries of the system, which is how supersymmetry is usually discussed in the context of
both classical and quantum field theories, though this structure is implicit in the way we constructed
the system from a gas of bosons and fermions.
We can take this winding rule as a new definition characterizing supersymmetric systems (at least free ones), 
even at finite temperature.

%This is most easily seen by using the standard form of the grand canonical free energy for bosons/fermions %and using
%L'H\^{o}pital's rule.)

Some open questions to ponder are: $(1)$ Is it possible to relate the sum over odd windings to a symmetry algebra?
It is evident from our construction that there is a symmetry relating bosons to fermions.
$(2)$ How do we include interactions into this picture? After all, the power of supersymmetry lies in its ability to help
us understand complicated interacting systems; free systems are too easy. Here perturbation theory, perhaps in the relativistic
notation we introduced by Feynman \cite{Feynman:1942us} should be analyzed.
$(3)$ Taking this new view, what do we learn about supersymmetric systems at finite temperature?
A well known property of supersymmetric systems is the non-renormalization theorems which protect certain quantities
against quantum corrections. These theorems are usually presented in the context of zero temperature quantum field theory.
Since, in the path integral formulation of the quantum statistical partition function, quantum fluctuations are associated with imaginary time dynamics, we conjecture that the contribution
made to any thermodynamic quantity by these fluctuations cancel, at all temperatures,
though at finite temperature, corrections arising from thermal fluctuations survive, a result
of the fact that away from $T=0$ the Bose and Fermi distributions differ from each other.
% and hence so do the average density of bosons or fermions feeding the density matrix.
It is usually claimed that supersymmetry is broken at finite temperature as a result of this
difference. For example, the mass renormalizations for bosons and fermions will be different.
Our argument suggests that their quantum renormalization still cancel, and the breaking
of the Bose/Fermi degeneracy is strictly a thermal effect.

\section{The enlightenment: Ceperley's constrained path integral.}

After these preliminaries, we have arrived at the core of this paper: Ceperley's 1991 discovery of a path integral representation for arbitrary fermion
problems that is not suffering from the 'negative probabilities' of the standard formulation\cite{Cepstatphys}. Surely, one cannot negotiate with the NP-hardness of the
fermion problem and Ceperley's path integral is not solving this problem in a mathematical sense. However, the negative signs are transformed away at the expense
of a structure of constraints limiting the Boltzmannian sum over world-line configurations. These constraints in turn can be related to a geometrical
manifold embedded in configuration space: the 'reach', which is a generalization of the nodal hypersurface characterizing wave functions to the fermion density matrix.
This reach should be computed self-consistently: it is governed by the constrained path integral that needs itself the reach to be computed. This is again
a NP-hard problem and Ceperley's path integral is therefore not solving the sign problem. However,  the reach contains all the data associated 
with the differences between bosonic and fermionic matter, and only its average and global  properties should matter for the physics in the scaling limit since it acts
on worldline configurations that themselves are averaged. Henceforth, it should be possible in principle to classify all forms of fermionic matter in
a phenomenological way by classifying the average geometrical- and topological properties of the reach, to subsequently use this data
as an input to solve the resulting bosonic path integral problem. This procedure is supposedly a unique extension of the Ginzburg-Landau-Wilson
paradigm for bosonic matter to fermionic matter. We do not have a mathematical proof that this procedure will yield a complete classification of fermionic
matter, but we have some very strong circumferential evidences in the offering that  it will work. The status of our claim is conjectural in the mathematical
sense.

Let us start out presenting the answer. Ceperley proved in 1991 that the following path integral is strictly equivalent to the standard fermion path integral
Eq. (\ref{pathdist},\ref{pathstat}),
        
\begin{equation}
\rho_F (\bR,\bR;\beta )=\frac{1}{N!}\sum_{{\cal P}, even}\int_{\gamma: \bR \to{\cal P}\bR}^{\gamma \in \Gamma_\beta(\bR)}
{\cal D}\bR e^{-\mathcal{S}[\bR]/\hbar}.  
\label{constpath} 
\end{equation}

This is quite like the standard path integral, except that one should only sum over {\em even} permutations (the reason to address this in section IV),  
while the allowed worldline configurations $\gamma$ are constrained to lie 'within the reach $\Gamma$'. This reach is defined as,

\begin{equation}
\Gamma_\beta(\bR) = \{ \gamma : \bR \rightarrow \bR' | \rho_F (\bR, \bR (\tau); \tau ) \neq 0 \}
\label{reach}
\end{equation}
for all imaginary times $0 < \tau < \hbar \beta$. In words, only those wordline configurations should be taken into account in Eq. (\ref{constpath}) that do not cause a 
sign change of the full density matrix at every intermediate imaginary time between $0$ and $\hbar \beta$.  In outline, the proof of this result is as 
follows. The fermion density matrix is defined as a solution to the Bloch equation

\begin{equation}
\frac{\ud \rho_F(\bR_0,\bR;\beta)}{\ud\beta}=-H \rho_F(\bR_0,\bR;\beta)
\label{Bloch}
\end{equation}
with initial conditions 

\begin{equation}
\rho_F(\bR_0,\bR;\beta=0)=\frac{1}{N!}\sum_{\cal P}(-1)^p\delta(\bR_0-{\cal P}\bR).
\label{initial}
\end{equation}
In the following we fix the reference point $\bR_0$ and define the reach $\Gamma(\bR_0,\tau)$ as before as the set of points $\{\bR_\tau\}$ for which there exists a continuous
space-time path with $\rho_F(\bR_0,\bR_{\tau'};\tau')>0$ for $0\leq \tau'<\tau$. Suppose that the reach is known in advance. It is a simple matter to show that the problematical initial 
condition, Eq. (\ref{initial}), imposing the anti-symmetry can be replaced by a zero boundary condition on the surface of the reach. It follows because the fermion density matrix 
is a unique solution to the Bloch equation (\ref{Bloch}) with the zero boundary condition. One can now find  a path integral solution without the minus signs. One simply restricts the paths to lie in the reach $\Gamma(\bR_0,\tau)$ imposing the zero boundary condition on the surface of the reach. The odd permutations fall for sure out of the reach since $\rho_F(\bR_0,{\cal P}_{odd}\bR_0)=-\rho_F(\bR_0,\bR_0)$.

The Ceperley path integral revolves around the reach. How to think about this object? The way the path integral is constructed seems to break imaginary time
translations. One has to first pick some 'reference point' ${\bf R}$ in configuration space at imaginary time $0$ or $\hbar \beta$. Starting from this set of
particle coordinates, one has to spread them out in the form of wordline configurations to check at every time slice that the density matrix does not change
sign. The dimensionality of the density matrix is $2dN + 1$ (twice configuration space plus a time axis) and the dimensionality of the reach is therefore
$2dN$  (one overall constraint). However, when we first pick a reference point ${\bf R}$ and we focus on a particular imaginary time the dimensionality
of this restricted reach is $dN-1$. In the limit $\tau \rightarrow \infty$  this restricted reach turns into a more familiar object: the nodal hypersurface associated
with the ground state wave function. The density matrix becomes for a given ${\bf R}$ in this limit, 

\begin{equation}
\rho({\bf R}, {\bf R'}; \beta = \infty) = \Psi^* ({\bf R}) \Psi ({\bf R'})
\label{grstden}
\end{equation}
and the zero's of the density matrix are just coincident with the nodes of the ground-state wave function, $\Psi ( {\bf R} ) = 0$, where we have assumed that the ground state is
non-degenerate. The wave function is anti-symmetric in terms of the fermion coordinates,

\begin{equation}
\Psi (\cdots, {\bf r}_i, \cdots, {\bf r}_j, \cdots) = - \Psi ( \cdots, {\bf r}_j, \cdots, {\bf r}_i, \cdots ),
\label{asymwf}
\end{equation}
and therefore the nodal hypersurface 

\begin{equation}
\Omega = \{ {\bf R} \in \mathbb{R}^{Nd} | \Psi ( {\bf R} ) = 0 \}
\label{nodalsurdeff}
\end{equation}
is a manifold of dimensionality $\textrm{dim}\Omega=Nd - 1$ embedded in $Nd$-dimensional configuration space.  This nodal surface $\Omega$ is surely an object that is 
simpler than the full reach $\Gamma$ and it is rather natural to train the intuition using the former. According to Ceperley's numerical results\cite{Cepstatphys}, it appears that at least 
for the Fermi gas   the main features of the reach are already encoded
in $\Omega$.  In a way, the dependence on imaginary time is remarkably smooth and unspectacular. A greater concern is the role of the reference point, or either the
fact that the reach depends on two configuration space coordinates. In the long imaginary time limit, the reach factorizes in the nodal surfaces (Eq. (\ref{grstden})),
which means that one can get away just considering the nodal surface of the ground state wave function, but this is not the case at finite imaginary times.  It is not at all that clear what role the
'relative distance' ${\bf R} - {\bf R'}$ plays, although there is some evidence that it can be quite important as we will discuss in Section IX.  Notice that the conventional
'fixed-node' quantum Monte-Carlo methods aim at a description of the ground state, using typically diffusion Monte-Carlo methods. As input for the 'fermionic-side',
these only require the wave function nodal structure. The difference between the reach and this nodal structure  is telling us eventually about
the special nature of the excitations in the fermion systems since the Ceperley path integral can be used to calculate dynamics, either in the form of finite temperature
thermodynamics or, by Wick rotation to real time, about dynamical linear response. At this moment in time it is not well understood what the precise meaning is of these
'dynamical signs' encoded in the non-local nature of the reach.  

\begin{figure}
\begin{centering}
\includegraphics[width=0.85\linewidth]{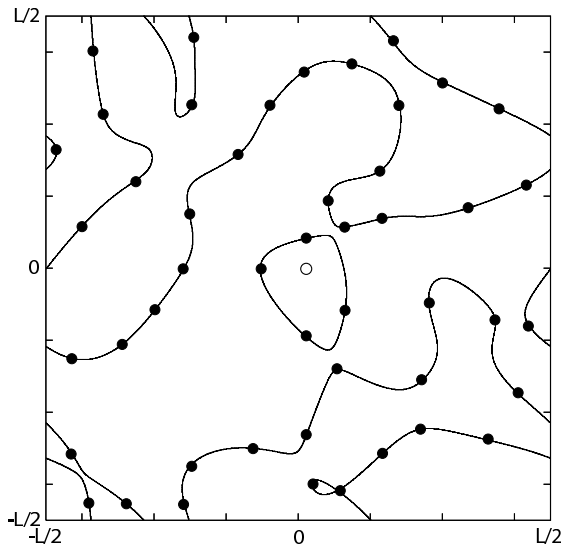}
\end{centering}
\caption{Cut through the nodal hypersurface of the ground-state wave function of $N=49$ free, spinless fermions in a two-dimensional box
with periodic boundary conditions. The cut is obtained by fixing $N-1$ fermions at random positions (black dots) and moving the remaining 
particle (white dot) over the system. The lines indicate the zeros of the wave function (nodes). Note that the nodal surface cut has to connect 
the $N-1$ fixed particles since the Pauli surface is a lower dimensional submanifold of dimension $Nd-d$ included in the nodal hypersurface
 with dimension $Nd-1$.}
\end{figure}

\begin{figure}
\begin{centering}
\includegraphics[width=0.75\linewidth]{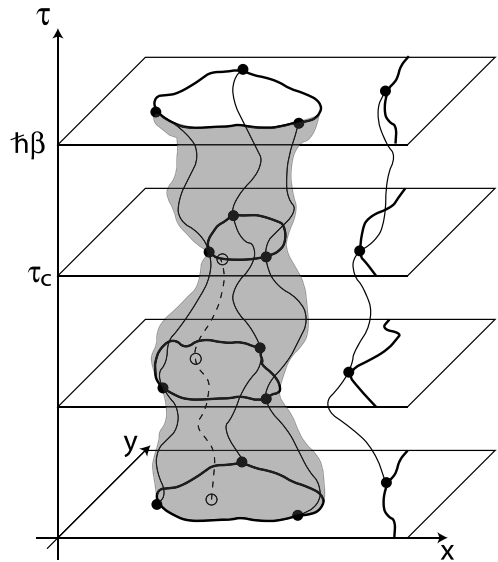}
\end{centering}
\caption{Nodal constraint structure in space-time seen by one particular particle. In the constraint path integral only world-line configurations $\{\bR_\tau\}$ are
allowed that do not cross or touch a node of the density matrix on all time slices, $\rho_F(\bR_0,\bR_\tau,\tau)\neq 0$ for $0\leq\tau<\hbar\beta$. Therefore, 
a particular particle (white circle) is constrained by the dynamical nodal tent (grey surface) spanned by the $N-1$ remaining particles trajectories (black circles). 
In a Fermi liquid the nodal tent has a characteristic dimensions and particles feel the nodal constraints at an average time scale $\tau_c$. In later chapter we will 
see that these scales are in one-to-one correspondence with the Fermi degeneracy scale $E_F$.}
\end{figure}

Another useful  geometrical object associated with Fermi-Dirac statistics is the Pauli surface, corresponding with the hypersurface in configuration space 
where the wave function vanishes because the fermions are coincident in real space,

 \begin{eqnarray}
 P &  =  & \bigcup_{i \neq j} P_{ij} \nonumber \\
 P_{ij} & = &  \{ {\bf R} \in \mathbb{R}^{Nd} | {\bf r}_i = {\bf  r}_j \}.
\label{paulisurdeff}
\end{eqnarray}

 Obviously, the Pauli surface is a submanifold of the nodal hypersurface of dimension $\textrm{dim}P=Nd - d$. The specialty of one dimension is that the
 Pauli- and nodal hypersurfaces are coincident. This property that the nodes are 'attached' to the particles is the key to the special status of one
 dimensional physics as we will explain in detail in the next section.  
  
In the next sections we will discuss in more detail the few facts that are known about the reach and  nodal hypersurface geometry and topology. To complete the discussion
of the basic structure of the Ceperley Path Integral, let us once more emphasize that  according to its definition Eq. (\ref{constpath}) one still has to sum over
 {\em even} permutations in so far these do not violate the reach. As for the signful path integral, this translates via the sum over cycles into a sum over winding numbers that are 
now associated with triple exchanges of particles.  We explained already in detail in section IV that this has the peculiar consequence that it codes for
supersymmetry when one is dealing with the free quantum gas that just knows about the even permutation requirement. Because of the constraints, the 'particles' of the
Ceperley path integral are actually very strongly interacting and it is unclear to what extent this supersymmetry is of any relevance to the final solution. In fact, we do know
for the Fermi-gas that the combined effect of the constraints and the triple exchanges is to eventually give back a free gas with Fermi-Dirac statistics. As we discussed in
section IV, there is  a 'don't worry theorem' at work because the thermodynamics of the supersymmetric gas is quite similar to the Bose gas.  
 
 In conclusion, Ceperley has demonstrated that in principle fermion problems can be formulated in a probabilistic, Boltzmannian mathematical language, paying the prize
 of a far from trivial constraint structure that is a-priori not known while it cannot be exactly computed. Qualitatively, the reach is like the nodal structure of a
 wave function. It is obvious that the nodal structure codes for physics but this connection is largely unexplored, while the remainder of this
 paper is dedicated to the case that it is actually quite easy to make progress,  at least with regard to the Fermi-liquid.  
 One particular  property is so important that it should be already introduced here. Any wave function of a system of fermions
 has the anti-symmetry property Eq. (\ref{asymwf}) and naively one could interpret this as 'any physical system of fermions has its fermionic physics encoded in a $Nd-1$ dimensional
 nodal surface'. This is obviously not the case. It is easy to identify a variety of fermionic systems where many more nodes are present in the fermion wave function
 than are required to encode the physics. A first example are Mott-insulating antiferromagnets on bipartite lattices. Because the electrons are localized they become effectively
 distinguishable.  One can therefore transform away remnant signs in the Heisenberg spin problem by Marshall sign transformations: the bottom line is that such Mott-insulators
 can be handled by standard bosonic quantum Monte Carlo methods.  A next example is physics in one dimensions, as we will discuss in the next section, where again
 the fermion signs can be transformed away completely, in a way that can be neatly understood in terms of the topology of the nodal surface.  Nodal structure is therefore
 like a gauge field: it carries redundant information that is inconsequential for the physics. Nodal structure that is in this 'gauge volume' we call {\em reducible} nodal
 structure, while the 'gauge invariant' (physical) part of the nodal structure we call  {\em irreducible}, and as a first step one should always first isolate the true, 
 irreducible signs. 
   
\section{The Ceperley path integral in 1+1 dimensions.}

The physics of quantum matter in one space dimension can be regarded as completely understood\cite{giamarchi}.  The deep reason is that quantum statistics has no
physical meaning in 1+1D, and it is always possible to find a representation where the sign structure drops out completely. All signs are reducible in the 
 language of the previous paragraph. Accordingly, the quantum problem is equivalent to a statistical physics problem in 2 classical dimensions,
and it appears that the problem solving capacity of statistical physics has no limit in this dimension. The reader might be familiar with the standard
bosonization techniques. A most elementary example is the Jordan-Wigner transformation which is usually introduced  to demonstrate
that $S=1/2$ quantum spin chain problems are equivalent to interacting spinless fermion problems, with as special cases the transversal field Ising model
(equivalent to 2D Ising) mapping onto free Majorana fermions\cite{kogut}, and the XY spin chain being equivalent to just free Dirac fermions\cite{fradkin}. It is instructive to find out how this
is processed by the Ceperley path integral. On the one hand, although the canonical Jordan-Wigner and bosonization methods are of course correct, the way
they deal with the (anti)symmetry of the states in Hilbert space is somewhat implicit and in this regard a discussion in terms of Ceperley's reach is most informative.
The other side is about the powers of fermionization; the simple free spinless Fermi-gas becomes in the Ceperley path integral representation a very serious statistical
physics problem. It is difficult to imagine a harder 2D statistical physics problem:  it is the 'Pokrovsky-Talapov' problem\cite{pokr}  of
 fluctuating polymers interacting via purely steric constraints. These correspond with infinitely strong
delta function potentials (real, finite range interactions simplify the problem!) and accordingly everything is about entropic interactions and order-out-of-disorder 
physics. Remarkably, this problem can be solved in a few lines using canonical fermions. Although the Ceperley path integral has a much richer structure in 
higher dimensions it is surely the case that the higher dimensional Fermi-liquids have to know in one or the other way about this 'entropic dynamics'. To highlight
this aspect we will review here the one dimensional fermion story in a less familiar, radically statistical physics way\cite{za1d,zamu1d}.

Let us first focus on the workings of quantum statistics in 1+1D, using the Jordan-Wigner transformation as a template. Consider a chain
of interacting $s=1/2$ spins, described by $SU(2)$ operators, $\left[ S^{\alpha}, S^{\beta} \right] = i \varepsilon^{\alpha \beta \gamma} S^{\gamma}$. Spins live in simple tensor
product space. In condensed matter physics they describe electrons that through a Mott condition got localized and localized electrons are 'Boltzmannions', i.e. distinguishable
particles. The standard construction continues claiming that on every site there are two available states (spin-up and -down) and this is no different from spinless 
fermions leaving a site unoccupied or singly occupied. But the difference is clearly in the antisymmetry of the fermion-Hilbert space, as encoded in the anti-commutation
property of the fermion operators.   This problem can be dealt with by the Jordan-Wigner sign string that works by the virtue that in one space dimension a string can
see all the particles covered by itself between its two end points, 

\begin{subequations}
\begin{eqnarray}
c(n) & = & \left(\prod_{j<n}[-\sigma^z(j)]\right)\sigma^-(n),\\
c^\dagger(n) & = & \sigma^+(n) \left(\prod_{j<n}[-\sigma^z(j)]\right).
\end{eqnarray}
\label{jorwig1}
\end{subequations}
Here $c(n)$ and $c^\dagger(n)$ denote fermionic annihilation and creation operators on site $n$, respectively, and $\sigma^\pm=(\sigma^x\pm\sigma^y)/2$ with
$\sigma^\alpha=\frac{2}{\hbar}S^\alpha$ the conventional spin-1/2 Pauli operators.

Having these operator identities it becomes then trivial to rewrite the spin-Hamiltonian in terms of the fermion operators and one finds out in
no time that the sign strings cancel out, and one obtains a problem that is local in the fermions. 

A little miracle has happened: we started out with Boltmannions and by the magic of the above operator identities we find out that we might as well consider
these distinguishable particles as fermions. In fact, one has the free choice to invoke hard core bosons as well in the intermediate stage since these share the property
with spins $s=1/2$ and spinless fermions that one has two available states per site. 

This is surely correct but in the canonical language it just appears as a mathematical fact associated with operator identities. The Ceperley path integral is in
this regard more transparent. Let us consider the meaning of the reach of the one dimensional Fermi-gas. For a given reference point ${\bf R}_0$ and 
imaginary time  $\tau$ one can associate with the fermion density matrix $\rho_F ({\bf R}_0,{\bf R}; \tau)$ a $(Nd-1)$-dimensional nodal hypersurface. However, we know that 
the $(Nd-d)$-dimensional Pauli surface is a submanifold of the nodal hypersurface and for $d=1$ the Pauli- and the nodal hypersurfaces have the same 
dimensionality and they are therefore the same! This is nothing else than the well established wisdom that in one dimension the nodes of the wave functions are attached
to the particle positions, a fact that is at the heart of Jordan-Wigner and all other bosonization constructions. In dimensions larger than one 'signs can have a life
of their own' because the nodal hypersurface has a larger dimensions than the Pauli surface. This is the simple but deep reason for the complete failure of
all attempts to construct Jordan-Wigner style bosonization procedures in higher dimensions.

Given that the Pauli- and nodal hypersurfaces coincide it becomes quite easy to read the reach. Start out with a reference point ${\bf R}_0 = ( x_1, x_2, \cdots, x_N)$
ordering the particles for instance like $x_1 < x_2 < \cdots < x_N$. 'Spread out' this configuration in terms of  world lines meandering along the time direction and the Pauli-hypersurface reach tells that only configurations are allowed where these worldlines never cross each other at any imaginary time. This is just the problem of an
ensemble of polymers with only steric, hard core interactions in 2 dimensions!  What is the fate of the quantum statistics? Let us permute two coordinates in the
reference point ${\bf R'}_0 = (x_2, x_1, \cdots, x_N)$. Because the particles one and two cannot pass each other these two starting configurations are disconnected:
they belong to two different nodal cells. Since this is true for any of the $N!$ permutations, in one dimensions one finds $N!$ nodal cells, instead of the two nodal
cells of the higher dimensional Fermi-gas, as discussed in the next section. The full partition sum consists of $N!$ copies of the same one-cycle 'Boltzmannion' partition function
starting from some particle sequence that is just divided by $N!$. The bottom line is that bosonic symmetry or fermionic anti-symmetry turns in the presence of the
Pauli-hypersurface reach  into a mere redundancy of the description. It has the status of a gauge volume and gauge invariant reality is caught in terms of a
Boltzmannion 'gauge fix'. Surely these wisdoms are well known from general considerations invoking the braid group, but the merit of the Ceperley path integral
is that it incorporates these considerations in a most explicit way.

\begin{figure}
\begin{centering}
\includegraphics[width=0.8\linewidth]{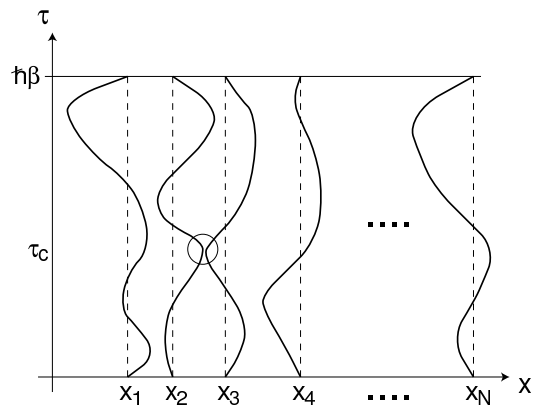}
\end{centering}
\caption{Allowed world line configurations in Ceperley path integral for one-dimensional fermions. In 1D the nodal hypersurface coincides with the Pauli surface and
the constraints turn into hard-core repulsions between the particles. Due to the hard-core constraints particle exchanges  are strictly forbidden and the particles become 
distinguishable. When the mean square displacement of the particles becomes of the order of the average inter-particle spacing, collisions between the particles start to happen.
This characteristic time scale $\tau_c$ is directly related to the Fermi energy, $E_F=\hbar/\tau_c$. From the statistical physics point of view the Fermi-gas in 1+1D just corresponds
with a problem of 'one-cycle' ring polymers interacting merely through steric constraints.}
\label{fig.1d}
\end{figure}

We now have landed on the statistical physics side: the Fermi-gas in 1+1D just corresponds with a problem of 'one cycle' ring polymers interacting merely
through steric constraints\cite{pokr}. Like the van der Waals liquid, problems with just steric interactions have a bad reputation. This is obviously about strong interactions
with the extra difficulty that the potentials are singular: a crossing of worldlines costs an infinite potential energy. The ramification is that all the physics
is driven by entropy. Surely, the easy way to solve this problem is to fermionize it and everything follows from the simple Fermi-gas solution. But how to read
these Fermi-gas wisdoms in the physical problem of the ring polymers? In fact, one of us\cite{za1d} was facing some time ago a problem which is quite similar to the 2D
ring polymers but where fermionization technology fails: the gas of hard core directed quantum strings in 2+1 dimensions, which is equivalent to the problem
of directed elastic membranes in 3D interacting through steric hindrance. This problem came up as a toy exercise in the investigation of quantum stripes in cuprates
and it is obviously the direct generalization of the 1+1D Fermi-gas to 2 space dimensions: just attach an extra space dimension and the world-lines spread out
in the string worldsheets. As it turns out, both the 1+1D Fermi gas and the 2+1D string gas can be addressed using a 'self-consistent phonon' method discovered 
by Helfrich\cite{helfrich} to deal with the entropic interactions associated with biological (extrinsic curvature) membranes. Let us just sketch the derivation.

In the polymer language, the partition sum of the Fermi-gas can be written as,

\begin{eqnarray}
\mathcal{Z} & = & \prod_{i=1}^N \prod_{\tau} \int d\phi_i (\tau) e^{\frac{1}{\hbar} \mathcal{S} }, \nonumber \\
\mathcal{S}  & = & \int d \tau \sum_i \frac{M}{2} ( \partial_{\tau} \phi_i )^2,
\label{onedpol}
\end{eqnarray}
where $\phi_i$ is the spatial displacement field of the $i$-th particle and the wordline is a 1D elastic manifold with
spring constant set by the mass $M$ of the particles. This is supplemented by the avoidance condition,

\begin{equation}
\phi_1 < \phi_2 < \cdots < \phi_N.
\label{avoid}
\end{equation}

Despite its simple formulation this is a rich problem characterized by various scales. Let us first inspect the ultraviolet  of the problem. The average distance 
between worldlines is $r_s =  L/N  = 1 / n$ where $L$ is the length of the system and $n$ the particle density. The worldlines only know about each others existence
when they collide because of the steric nature of the interactions. Henceforth, at sufficiently short times the worldlines will have meandered over distances that are
small compared to $r_s$ and they cannot have knowledge of each others existence. The problem becomes cooperative at the characteristic time scale for  collisions
to occur.  This is straightforward to estimate. By detailed balance the mean square fluctuation of the elastic line grows with $\tau$ as,

\begin{equation}
l_c^2 (\tau) = \langle ( \phi_i (\tau) - \phi_i (0) )^2 \rangle = \frac {\hbar} {M} \tau.
\label{msqfluc1D}
\end{equation}
When $l_c (\tau)$ becomes similar to $r_s$ a collisions will occur and this will take a typical time

\begin{equation}
\tau_c \simeq \frac{M}{\hbar} n^{-2},
\label{coltime1D}
\end{equation}
which in the quantum interpretation this time is associated with an energy,

\begin{equation}
E_F = \frac{\hbar}{\tau_c} = \frac{\hbar^2} {2m} n^2,
\label{colener1D}
\end{equation}
and we recognize the Fermi energy\cite{za1d}! This makes perfect sense. The Fermi energy is the characteristic energy where the effects of the quantum statistics on 
the otherwise free system becomes  noticeable. In the Ceperley path integral this translates into the time scale where for the first time the particles
become aware of the presence of the constraints. In the next section we will find that the above argument can be trivially extended to higher dimensions.

The Fermi energy is easy but now the trouble starts. When the collisions start to happen the problem becomes highly cooperative and in principle hard to deal
with in terms of exact methods. The key is entropic repulsions, the same effect causing a rubber band to stiffen up when you heat it. The qualitative argument
runs as follows\cite{za1d}. Take the gas of non-interacting polymers as a reference point. In this free gas there are two possibilities when two polymers meet each other:
either they cross or do not cross. But in the steric avoidance gas, there is only one possibility (do not cross) and therefore the system has to pay an entropy
cost of $\log (2)$ at every collision relative to the non-interacting system. This adds a positive (repulsive) term $ n_c (T)  k_B T \log (2) $ to the free energy
where $n_c(T)$ is the density of collisions at temperature $T$. This will have the effect that upon coarse graining the polymers start to repel each other:
the collisions at short distances caused by the strong microscopic fluctuations renormalize at large distances into 'entropic springs' keeping the polymers apart!

To make this more quantitative\cite{za1d} we need the ingenious trick devised by Helfrich\cite{helfrich}. Quite generally, the effect of the entropic repulsions will be to build up crystalline
correlations and in the scaling limit one has to find a crystal with algebraic long range order (2D, finite temperature or 1+1D at finite coupling constant). This 
worldline crystal does not carry shear rigidity because the worldlines are incompressible in the time direction. Henceforth, the crystal is characterized by a space-direction 
compressional modulus $B_0$, besides the time direction mass 'spring constant'. The coarse-grained action can be written in terms of the effective elastic fields $\psi$ as

\begin{equation}
S_\textrm{eff} = \frac{1}{2} \int d\tau \int d x \left[  \rho ( \partial_{\tau} \psi )^2 + B_0 (\partial_x \psi)^2 \right],
\label{heleffaction}
\end{equation}
where $\rho = n M$ is the mass density. Now Helfrich's trick comes: for finite $B_0$ fluctuations are suppressed  relative to the case that $B_0$ vanishes and this
entropy cost raises the free energy by an amount $\Delta F (B_0) = F(B_0) - F(B_0 = 0)$. But by general principle it has to be that the 'true' long wavelength 
modulus $B$ should satisfy

\begin{equation}
B =   r_s^2 \frac{\partial^2} {\partial r_s^2} \left(  \frac {\Delta F (B_0)} {L} \right),
\label{helfrich}
\end{equation}
and in the case of steric interactions, the only source of long wavelength rigidity is the fluctuation contribution to $\Delta F$. This implies that $B_0 = B$ and 
Eq. (\ref{helfrich}) turns into a self-consistency condition for this 'true' entropic modulus $B$!  It is an easy exercise to work this out and one finds 

\begin{equation}
B= \frac{9 \pi^2} {\eta^4} \frac{ \hbar^2} {M r_s^3},
\label{renormmod}
\end{equation}
where $\eta$ is a fudge factor associated with the ultraviolet cut-off; the self-consistent phonon theory should set in at a length $x_\textrm{min}$ where crystalline correlations
become noticeable and because this requires some number of collisions $x_\textrm{min} = \eta r_s$ where $\eta > 1$. 

To identify this 'order out of disorder' physics with our canonical Fermi-gas we have to invoke some more bosonization wisdoms. The bottom line of the Tomonaga-Luttinger liquid (Luttinger liquids are connected adiabatically to the Fermi-gas) is that he electron system is nothing else than a 1+1D 'floating' (algebraically ordered)
crystal, characterized by a spectrum of compressional phonons, the bosonization modes. This spectrum is characterized by a single dimension: the sound
velocity that is coincident with the Fermi velocity in case of the free Fermi-gas. This velocity is in  the 'entropic world' Eq. (\ref{heleffaction}) given by
 $v_F = \sqrt{ B / \rho}$ and the correct answer follows from Eq. (\ref{renormmod}) for a reasonable $\eta = \sqrt{6}$. 
 
 To complete this story, one can treat the 2+1D directed string gas with the same methodology\cite{za1d} to find out the peculiar result that now both the Fermi
 energy and the Fermi-velocity become exponentially small in the density, like $v_F \sim \exp{(- const/ \mu^{1/3}) }$ where $\mu = \hbar / (\rho c d^2)$ with $c$
 the worldsheet velocity and $d$ the interstring distance. This reflects the fact that strings fluctuate a lot less than particles, and the entropic interactions are
 suppressed. But in any other regard this string gas is quite like the 1+1D Fermi gas! 
 
 We hope that the reader has appreciated this story. We perceive it to an extent as demystifying. The bottom line is that the rather abstract mathematical procedures 
 of one dimensional physics are just coding accurately for a physical world that is dominated by entropic interactions and order-out-of-disorder physics. It hits home
 the case that it is quite misleading to call the Fermi-gas a 'gas'. One could argue naively like: 'the system only knows about kinetic energy so how can it be anything
 else than a gas'? But Fermi-Dirac statistics translates into a statistical physics via Ceperley's path integral characterized by a structure of constraints that is essentially 
 steric in nature. This is obviously the case in 1+1D but  in this basic regard things are the same in the less well understood higher dimensional cases. These 
 steric constraint problems have as a generic feature that the microscopic 'kinetic energy' fluctuations and the macroscopic 'potential energy' scales associated with
 order are governed by the same dimensions. Order-out-of-disorder is in these kind of worlds an ubiquitous, hard to avoid mechanism and the bottom line is that 
 all fermion systems in 1+1D fall victim to the algebraic order. This is in turn the key to the 'universal success' of bosonization\cite{giamarchi}. Bosonization is just geared to deal accurately
 with the fluctuations around the ordered state. Turning to the spin-full systems there is more life than just crystallization (and umklapp pinning).  It becomes possible for
 wordlines to 'come to an end' and these correspond with dislocations in the space time worldline crystals that are dynamically indistinguishable from XY vortices.
 Henceforth, one finds a quantum melting equivalent to the 2D Kosterlitz-Thouless transition leading into the gapped Luther-Emery state\cite{luem}. Another fascinating
 feature is the hidden 'squeezed lattice' geometrical order that lies at the heart of spin-charge separation. This is accurately encoded in standard bosonization,
 but only recently characterized in full using a statistical physics style non-local order parameter structure\cite{kruis}.
 
 There is much more to tell about physics in one dimensions, but this is less fun because it is fully understood. The take home message is a warning: even when
 one fully grasps the nodal structure (not true in higher dimension), while the problem maps onto something that exists in classical nature (not true either), a seemingly
 trivial fermion problem (the 1+1D Fermi gas) turns out to code for a remarkably complex and rich world devoid of fermion signs!

\section{The Fermi gas as a cold atom Mott-insulator in momentum space.}

\begin{figure}
\begin{centering}
\includegraphics[width=0.95\linewidth]{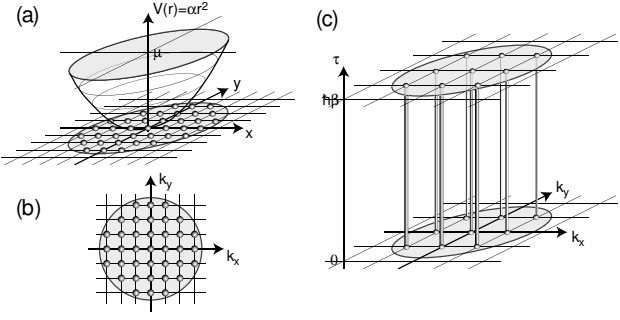}
\end{centering}
\caption{(a) a system of {\em classical} atoms forming a {\em Mott insulating state} in the
 presence of a commensurate {\em optical lattice of infinite strength}, living in a  {\em harmonic potential trap $V(\vec{r})=\alpha r^2$} of finite strength; (b) the trap in momentum space $k_{x},k_y$ instead of real space; the Fermi surface is just the 
 boundary between the occupied optical lattice sites and the empty ones; (c) a grid of allowed momentum states $k = (2 \pi/L) (k_x, k_y, k_z, ....)$ where 
the $k_i$'s are the usual integers and any worldline just closes on itself along the imaginary time $\tau$ direction $0\rightarrow \beta$ : single particle momentum conservations prohibit anything but the one cycles.}
\label{fig.fermi}
\end{figure}

 The Fermi-gas of the canonical formalism is very easy to solve exactly, and one would expect that in one or the other
 way this should mean that the constrained path integral is also easy to solve. This is not true at all in the position representation,
 as we will discuss in the next section. However, considering the derivation of the Ceperley path integral there is actually no
 preferred status of real space. The construction is completely independent of the representation one chooses for the 
 single particle states. On the canonical side momentum space  is the convenient representation to start from in the galilean 
 continuum, or either any other basis that diagonalizes the single particle problem. As we will show in this section, also the
 Ceperley path integral of the Fermi-gas becomes very easy indeed when one chooses to formulate it in momentum space. After 
 a couple of straightforward manipulations one finds a sign free, Boltzmannian path integral showing a most entertaining correspondence:
 the Fermi-gas is in one-to-one correspondence with {\em a system of classical atoms forming a Mott insulating state in the
 presence of a commensurate optical lattice of infinite strength, living in a  harmonic potential trap} of finite strength (see Fig. \ref{fig.fermi}a). 
 This is literal and the only oddity is that this trap lives in momentum space instead of real space; the Fermi surface is just the 
 boundary between the occupied optical lattice sites and the empty ones. This boundary is sharp at zero  temperature but it smears at finite
 temperature because of the entropy that can be gained by exciting atoms out of the trap!  When you are quick, you should already have
 realized that this trap interpretation is actually consistent with everything we know about the Fermi-gas. Let us now proof it by
 constructing the Ceperley path integral.
 
 The central wheel of the Ceperley path integral is the fermion density matrix. One should first guess an ansatz, use it to construct the path integral, to
 check if the same density matrix is produced by the path integral.  Surely we know the full fermion density matrix for the Fermi gas, and in
 momentum space this turns out to be a remarkably simple affair. The k-space  density matrix can be written as the determinant formed from 
imaginary time single particle propagators in the galilean continuum,

\begin{equation}
g ({\bf k}, {\bf k}';\tau) = 2\pi\delta ( {\bf k} - {\bf k}' ) e^{  -\frac{|k|^2 \tau}{2\hbar M} }.
\label{kpropagators}
\end{equation}

Since we live in the space of exact single quantum numbers these propagators are diagonal; in the galilean continuum
this just means the conservation of momentum, but when translational symmetry is broken one should use
here just the basis diagonalizing the single particle Hamiltonian.

Consider now the full momentum configuration space ${\bf K}=(\mathbf{k}_1,\ldots,\mathbf{k}_N)$ imaginary time density matrix,

\begin{equation}
\rho_F ( {\bf K}, {\bf K}'; \tau ) = \frac{1}{N!} \sum_\mathcal{P} (-1)^p \prod^{N}_{i=1}  g\left({\bf k}_{p(i)},{\bf k}_{i}';\tau\right).
\label{Kdensmatrix}
\end{equation}

We find that the delta functions cause a great simplification. Substituting  the single-fermion expression  Eq. (\ref{kpropagators})    in  this expression for the density matrix Eq. (\ref{Kdensmatrix})  we obtain:

\begin{eqnarray}
\rho_F ( {\bf K}, {\bf K}'; \tau ) &  = & \frac{1}{N!} e^{-\sum^{N}_{i=1}\frac{|{\bf{k}}_{i}|^2 \tau}{2\hbar M} }\nonumber\\
& & \times \sum_\mathcal{P} (-1)^p  \prod^{N}_{i=1}2\pi\delta ( {\bf k}_{p(i)} - {\bf k}_{i}' ).\quad
\label{Kdensmatrix1}
\end{eqnarray}

Since the single particle propagators are eigenstates of the Hamiltonian, the momentum world lines go 'straight up' in the time direction until they arrive at the time $\tau$ where
the reconnections can take place associated with the permutations. But the $\delta$ function enforces that the permuted momentum has to be the same as the non-permuted
one, and the worldlines can therefore not wind except  when the momenta of some pairs of fermions coincide. But now the sum of the permutations in Eq. (\ref{Kdensmatrix1}) 
is zero due to the Pauli principle. Mathematically, this follows from the fact that the expression on the right hand side of Eq. (\ref{Kdensmatrix1}) is actually a Slater determinant 
formed from the delta-functions $2\pi\delta ( {\bf k}_{p(i)} - {\bf k}_{i}' )$ as the matrix elements of the $Nd\times Nd$ matrix, that are indexed by momenta $\{{\bf k}_{p(i)},{\bf k}_{i}'\}$.
Hence, when two of the momenta coincide (e.g. ${\bf k}_{i}={\bf k}_{j}$, $i\neq j$ )there are two coinciding raws/columns in the matrix and the Slater determinant equals zero.  
The result is that Eq. (\ref{Kdensmatrix}) factorizes in $N!$ relabeling copies, associated with $N!$ nodal cells like in 1+1D,  of the following simple density matrix describing distinguishable and localized particles in momentum space,

\begin{equation}
\rho_F ( {\bf K}, {\bf K'}; \tau ) =   \prod^{N}_{{\bf k}_1 \neq {\bf k}_2 \neq \cdots \neq {\bf k}_N} 2\pi\delta ( {\bf k}_{i} - {\bf k}_{i}' ) e^{-\frac{|{\bf{k}}_{i}|^2 \tau}{2\hbar M} }. 
\label{Kdensmatrixcep}
\end{equation} 

This has the structure of a Boltzmannian partition sum of a system subjected to steric constraints:
 it is actually the solution of the Ceperley path integral for the Fermi gas in momentum space!  Let us apply 
periodic  boundary conditions so that on every time slice of the Ceperley path integral we find a grid of allowed momentum states ${\bf k}_i = (2 \pi/L) (k_{i,x}, k_{i,y}, k_{i,z}, ....)$ where 
the $k_{i,\alpha}$'s are the usual integers (see Fig. \ref{fig.fermi}b). We learn directly from Eq. (\ref{Kdensmatrixcep}) that we can ascribe a distinguishable particle with every momentum cell,
with a worldline that just closes on itself along the time direction: single particle momentum conservation prohibits anything but the one cycles (see Fig. \ref{fig.fermi}c).
In addition,  we find that the reach just collapses to the Pauli hypersurface, just as in one dimensions: per momentum space cell either zero
or one worldline can be present. These worldlines are given by Eq. (\ref{kpropagators}): since we are living in exact quantum number space 
these just go straight up along the time direction, since there are no quantum fluctuations: these are actually classical particles living in
momentum space. We do have to remember that these world 'rods' carry a fugacity set by a potential   $ \frac{|k|^2 \tau}{\hbar M}$. Henceforth,
we have a problem of an ensemble of classical hard core particles that live on a lattice of  'cells' in momentum space where every cell can either
contain one or no particle, with an overall harmonic potential envelope centered at ${\bf k} = 0$: this is literally the problem of cold atoms living
in a harmonic trap, subjected to an infinite strong optical lattice potential, tuned such that they form a Mott-insulating state.  The ground state
is simple: occupy the cells starting at ${\bf k}=0$, while the particles are put into cells at increasing trap potential until the trap is filled up with the
available particles. At zero temperature there are no fluctuations and when one exceeds the chemical potential the cells remain empty, and there
is a sharp $(d-1)$-dimensional interface between the occupied- and unoccupied trap states. This is of course the way we explain the Fermi-gas to
our undergraduate students. It invokes an odd metaphor that however turns out to express an exact identification since we learned to
handle the Ceperley path integral!  
 
 Having a statistical physics interpretation, can we now address the questions posed in section II? First, what is the order parameter of the Fermi-liquid?
 The answer is: the same order parameter that governs the Mott-insulator. This order parameter is well understood\cite{fradkin}, although it is of an unconventional
 kind: it is the 'stay at home' emergent $U(1)$ gauge symmetry\cite{baska}, stating that at every site and at all times there is precisely one particle per site. The particle
 number is locally conserved and henceforth a local $U(1)$ symmetry emerges. The 'disorder operators' that govern the finite temperature fate of the order parameter are
 just substitutional-interstitial defects: there is a finite thermal probability to excite a particle out of the trap, and the presence of the vacancies
 destroys the $U(1)$ gauge symmetry.  Since the disorder operators are zero-dimensional particles regardless the dimensionality of momentum space, thermal
 melting of the Mott-insulator occurs at any finite temperature regardless dimensionality. In the next section we will discuss how this might relate to the "holography"
introduced in section II.
 
 We repeat, this is just a rephrasing of the standard Fermi gas wisdoms in a non-standard language. The strange powers of the Ceperley path integral become more
 obvious when interactions are switched on. In the presence of the interactions single-particle momentum is no longer conserved, and this means that
 the worldlines of the Ceperley particles in momentum states get quantized: it is analogous to making the optical potential barriers finite in the cold gas Mott-insulator
 with the effect that the particles acquire a finite tunneling rate between the potential wells. One gets directly a hint regarding the stability of the Fermi-liquid:
 Mott-insulators are stable states that need a rather large tunneling rate to get destroyed. But the story is quite a bit more interesting than that, as can be
 easily argued from the knowledge on the canonical side. Let's consider first what would happen in a literal cold atom Mott insulator when we start to
 quantize the atoms. Deep inside the trap motions are only possibly by doubly occupying the nodal cells and given that in the non-interacting limit the
 'Hubbard U' is infinite (expressing the Pauli surface) such processes are strongly suppressed. In the bulk of the trap the Mott state would be very robust.
 However, at the boundary one can make cheap particle-hole excitations, and at any finite $t$ the
 interface would no longer be infinitely sharp on the microscopic scale: the density profile would change smoothly. Eventually one would meet the 'wedding
 cake' situation where the bulk is still Mott-insulating while the interface would turn into a superfluid (we live in a bosonic world). How different is the
 Fermi-liquid!  We know how it behaves from the canonical side. The single-fermion self-energy tells us directly about the fate of the ${\bf k}$-space Mott insulator.
 We learn that the  time required to  loose information on single-particle momentum is just given by the imaginary part of the self-energy and that behaves
 as\cite{AGD} $1/ \tau_k \sim ( k-k_F)^2$, Henceforth, it diverges at the interface while it get shorter moving into the bulk. In the Ceperley bosonic language the 
 Fermi-liquid is like a grilled marshmallow: It has a 'crispy', solid Mott insulating crust while it  becomes increasingly fluid when one moves inside!
 
More precisely,  the worldlines near the interface are fluctuating at short times, since we know that the momentum distribution of the bare 
 electrons do smear around the Fermi-momentum - they do 'spill out of  the trap'. However, the effect of integrating out these microscopic fluctuations is
 to renormalize the 'optical lattice potential' upwards. This has to be the case because in the scaling limit the renormalized worldlines represent the
 quasiparticles and since they produce a perfectly sharp interface (i.e. unit jump in the quasiparticle $n_k$), the Mottness has to be perfect. This
 can only be caused by infinitely high effective potential barriers. This physics is of course coming from the modifications happening in the reach when 
 interactions are turned on. The phase space restrictions giving rise to $\Sigma''  \sim \omega^2$ are rooted in Fermi-Dirac statistics and all the statistical
 effects are coded in the reach when dealing with the Ceperley formalism. These aspects can be computed by controlled perturbation
 theory and in a future publication they will be analyzed in detail.     

 \section{The Fermi-liquid in real space: holographic duality.}
 
We showed in the previous section that at least for the Fermi gas the momentum space Ceperley path integral becomes
a quite simple affair. Momentum space is a natural place to be when one is dealing with a quantum gas or
-liquid, but dealing with a bosonic- or statistical physics systems one invariably runs into the general notion
of duality\cite{kogut,cvetkovic}. Dealing with conjugate degrees of freedom, like momentum and position or phase and number, one can reformulate
the manifestly local order on one 'side' into some non-local topological order parameter on the dual side. An
elementary example is the Bose-Einstein condensate. In the language of the previous section, one can either form a 'black
hole' in the momentum space 'trap', by putting all bosons in the ${\bf k} = 0$ 'optical lattice cell'. But one can also view it in real space, to 
discover the lively world of Section III where the local order in momentum space translates into a global, topological description revolving
around the infinite windings of worldlines around the time direction. Such duality structures are ubiquitous in Bolzmannian systems, and
they are at the heart of our complete understanding of such systems: when one has a complete duality 'map' one understands the system 
from all possible sides and there is no room for surprises. For instance, when one is dealing with a strongly interacting system like $^4$He one
prefers the real space side because it is much easier to track the effects of the interactions\cite{ceperhelium}. Also in the strongly interacting fermion systems one
expects that one is better off on the real space side.  In this concluding section we will address the issue of the dual, real space description
of the Fermi-liquid in the Ceperley path integral formalism. This real space side is remarkably complex: despite an intense effort even
Ceperley and coworkers\cite{cepprivat} got stuck to the degree that they even did not manage to get things working by brute computer force. They ran 
into a rather mysterious  'reference point glassification' problem in their quantum Monte Carlo simulations, likely related to a contrived
'energy landscape' problem associated with the workings of the reach.      

This is a fascinating problem: there has to be a simple, dual real space description of the Fermi gas.  The obvious difficulty as
compared to straightforward bosonic duality is the presence of the reach. One has to dualize not only the 'life of the worldlines' but
also the constraints coding for the Fermi-Dirac statistics. Topology is at the heart of duality constructions and in this regard
Ceperley\cite{Cepstatphys}, and more recently Mitas\cite{Mitas}, have obtained some remarkably deep results, which will be discussed at length in the first
subsection: the topology of the reach of the Fermi-liquid in $d \ge 2$ is such that the reach is open for all cycles of Ceperley worldlines
based on even permutations or triple exchange. Henceforth, there is no topological principle that prevents infinitely long
worldlines to occur and in subsection B we will  argue that the zero temperature order of the Fermi-liquid has to be a Bose condensate
of the 'Ceperley particles'.  This is conjectural but if it proves to be correct  the Fermi-liquid holography we discussed in section II acquires
a fascinating meaning:
{\em the scaling limit thermodynamics of the Fermi-gas in any spatial dimension $d > 1$ is governed entirely by the statistical
physics associated with distributing the Ceperley worldlines over the cycles associated with even permutations, and this 
effective partition sum is indistinguishable from the partition sum enumerating the cycles of a soft-core boson system in
one space dimension.}

 \subsection{The topology of the Fermi-liquid nodal surface.}    

To decipher the structure of constraints as needed for the real space Ceperley path integral one has to find out where the zero's of the 
real space density
matrix are. By continuation, these should be in qualitative  regards the same in the Fermi-liquid as in the Fermi gas, and in the latter case we have
an expression of the full density matrix in closed form,  

\begin{eqnarray}
\rho_F ({\bf R}_0, {\bf R}; \tau ) & = &  ( 4 \pi \lambda \tau)^{-dN/2} \nonumber\\
& & \times\det  \exp \left[ - \frac{ ( { \bf r}_i- { \bf r}_{j0} )^2 }{4 \lambda \tau} \right],
\label{densmat0}
\end{eqnarray}
where $\lambda = \hbar^2 / (2M)$. Henceforth, one needs to find out the zero's of this quantity for all ${\bf R}_0, {\bf R}$ in the imaginary
time interval $0 < \tau < \beta$.  In real space, this is not an easy task. Part of the trouble is that at low temperature the zero's of the determinant
depend on all coordinates at the same time. Only in the high temperature limit $(\tau\to 0)$ the nodal surface of the density matrix 
becomes extremely simple\cite{Cepstatphys}. To see this, define first a {\em permutation cell} $\Delta_\mathcal{P}({\bf R}_0)$ as the set of points closer to
$\mathcal{P}{\bf R}_0$ than to any other $\mathcal{P}'{\bf R}_0$. Obviously, the configuration space is divided into $N!$ permutation cells which are
convex polyhedra bounded by hyperplanes, ${\bf R}\cdot(\mathcal{P}{\bf R}_0-\mathcal{P}'{\bf R}_0)=0$. The density matrix is simply a sum
over all permutations and for ${\bf R}\in\Delta_\mathcal{P}({\bf R}_0)$ and sufficiently high temperatures this sum is completely dominated by 
the term $(-1)^p\exp[-({\bf R}-\mathcal{P}{\bf R}_0)^2/(4 \lambda \tau)]$ since all the other terms are exponentially damped relative to it. Therefore, 
in the high temperature limit, $\rho_F ({\bf R}_0, {\bf R}; \tau)$ will have the sign of $\mathcal{P}$ inside of $\Delta_\mathcal{P}({\bf R}_0)$ and the nodal 
hypersurface is simply given by the common faces shared by permutation cells of different parities.

The reach acts both in a local way, much in the same way as we learned in the (1+1)-dimensional case as a special 'steric hindrance' structure having
to do with entropic interactions, etcetera. However, it also carries global, topological properties and these are now well understood 
because of some remarkable results by Mitas\cite{Mitas}, who managed to proof the 'two nodal cell' (or 'nodal domain') property of the higher dimensional 
Fermi-gas reach\cite{Cepstatphys}. The topology of the nodal surface is associated with the structure of cycles as discussed  in section III but now for the
Ceperley path integral. The latter can be written as

\begin{equation}
Z = \sum_{\mathcal{P}_{e}} \int d {\bf R} \tilde{\rho}_D ( {\bf R}, \mathcal{P}_e {\bf R}; \beta ),
\label{cepper}
\end{equation}
where $\mathcal{P}_e$ refers to even permutations, while $\tilde{\rho}_D$ refers to the density matrix of distinguishable particles that are however still
subjected to the reach constraints.  As in the case of the Feynman path integral, this sum over even permutations can be recasted in a sum over cycles
associated with all possible ways one can reconnect the worldlines at the temporal boundary, of course limiting this sum to those cycles that
are associated with even permutations.  We learned in section IV that for free wordlines even permutations translate into the supersymmetric quantum
gas. But the Ceperley particles are not at all free, and the topology of the nodal surface tells us about global restrictions on the cycles that can 
contribute to Eq. (\ref{cepper}).

It is immediately clear that the counting of cycles is governed by topology: to find out how to reconnect wordlines arriving at the temporal boundary 
from the imaginary time past, to worldlines that depart to the imaginary time future one needs obviously {\em global} data. This
global information residing in the reach is just the division of the reach in nodal cells we already encountered in the (1+1)-dimensional context and  the
momentum space Fermi gas. There
we found that the space of all permutations got divided in $N!$ nodal cells, with the ramification that  the sum in Eq. (\ref{cepper}) is actually reduced
to one cycles. Mitas has delivered the proof that in $d\ge 2$ the reach carries a two nodal cell topology, implying
that {\em all cycles based on even permutations lie within the reach.} Since only this topological property of the reach  can impose that certain cycles 
have to rigorously disappear from the cycle sum, this does imply that all cycles based on even permutations can contribute to the partition sum, including
the cycles containing macroscopic winding  numbers. Henceforth, the Ceperley worldlines can Bose condense in principle and it is now just matter
of finding out what the distributions of the winding numbers are as function of temperature. This is what really matters for the main line of this story.
Finding out the the way that Mitas determined the two-cell property is quite interesting and we will sketch it here for those who are interested. When
you just want to understand the big picture, you might want to skip the remainder of this subsection.  

 Quite recently Mitas\cite{Mitas} proved a conjecture due to Ceperley\cite{Cepstatphys}, stating that the reach of the higher dimensional Fermi gas is 'maximal' 
 in the sense that, for a given ${\bf R}_0$ and $\tau$, the nodal surface of $\rho_{F} ({\bf R}_0, {\bf R}; \tau )$ separates the configuration space in just two nodal cells, corresponding with $\rho_{F} $ being positive- and negative respectively. This is a quite remarkable property:
 for every pair ${\bf R}$ and ${\bf R}'$ in the same domain (lets say $\rho_F > 0$), one can change ${\bf R}$ into ${\bf R}'$ without encountering
 a zero crossing of $\rho_F$.  
 
 The easy way to prove this property goes as follows\cite{Mitas}. First, it can be demonstrated\cite{Cepstatphys} that once there are only two nodal cells at some
 initial $\tau_0$ than this property has to hold for any $\tau > \tau_0$. This follows straightforwardly from the imaginary time Bloch equation for the density
 matrix,
 
 \begin{equation}
 - \frac{ \partial \rho ( {\bf R}, {\bf R'}; \tau )}{\partial \tau} = H \rho  ( {\bf R}, {\bf R'}; \tau )
 \label{blocheq}
 \end{equation}
 with initial condition,
 
 \begin{equation}
 \rho ( {\bf R}, {\bf R'}; 0 ) = \det \left[ \delta ( {\bf r}_i - {\bf r}'_j ) \right]
 \label{blochbound}
 \end{equation}
and the Bloch equation is a linear equation.  This is
a very powerful result because it gives away that the two-cell property 'descents for the ultraviolet': one has just to prove it at an arbitrary short imaginary
time which is the same as arbitrary high temperature. Ignoring Planck scale uncertainties, etcetera, the form Eq. (\ref{densmat0}) has to become asymptotically
exact for sufficiently small $\beta$, also in the presence of arbitrary interactions as long as they are not UV-singular! As we already noticed, this high temperature
limit is rather tractable.

We now need to realize that we still have to take into account the 'remnant' of quantum statistics in the form of even permutations. Every even permutation can
be written as a succession of exchanges of three particles $i, j, k \rightarrow j, k, i$ because these amount to two particle exchanges. When such an exchange
does not cross a node (i.e. it resides inside the reach) the three particles are called 'connected'. By successions of three particle exchanges one can build up
clusters of connected particles. All one has now to demonstrate is that a point ${\bf R}_t$ exists where {\em all} particles are connected in a single cluster, because
this complete set of even permutations exhaust all permutations for a cell of one sign, because the odd permutations necessarily change the sign. One now needs
a second property called tiling stating that when the particles are connected for the special point  ${\bf R}_t$ this has also to be the case for all points in the cell.
And tiling is proved by Ceperley for non-degenerate ground states and also for finite temperature. Actually due to the linearity of the Bloch equation, its fixed node solution is unique, and the tiling property in the high temperature limit will lead to the same property at any lower temperature.  

Before we prove that the above holds for the high temperature limit density matrix, let us just dwell for a second on what this means for the winding properties
of the constrained path integral. The even permutation requirement means that, as for the standard worldline pathintegrals, we have to connect the worldlines
with each other  at the temporal boundary, but now we have to take care that we single out those cycles corresponding with even (or three particle) exchanges.
The 'maximal reach' just means that cycles containing worldlines that wind an arbitrary large number of times around the time axis {\em never encounter a node !}
As noted before by Ceperley, this has the peculiar implication that in some non-obvious way the Fermi-gas has to know about Bose condensation. Since nodal
constraints do allow for infinite windings there seems to be no 'force in the universe' that can forbid these infinite windings to happen and since the Cepereley
path integral is probabilistic, when these infinite windings happen one has to accept it as Bose condensation. We will come back to this theme in a moment.

Following Mitas, one can now prove the two cell property of the high temperature limit using an inductive method. Assume that all $N$ particles in the low
$\beta$ limit at a fixed ${\bf R_0}$ are connected in one cluster, to see what happens when an additional $N+1$ particle is added. Single out two other particles
$N-1$, $N$ and move these three particles away from the rest without crossing a node. Now we can profit from the fact that in the low $\beta$ limit the density
becomes factorizable: the determinant factors into a product of the determinant of the three special particles and the determinant of the rest. It is easy to show
that the three particle determinant has the two cell property, proving that the N+1's particle is in the cluster of $N$ particles. Since this is true for any $N$, the
starting assumption that all particles in the cluster is hereby proven.

For free fermions, Mitas also proved the two nodal cell property for non-degenerate ground states using a similar induction procedure. The trick is to choose a special point ${\bf R}_t$ in the configuration space, at which one can easily show how all the particles are connected into a single cluster. Once proven for this single point, tiling ensures that the same is true for the entire nodal cell. Mitas aligned the particles into lines and planes, thus forming some square lattice in the real space. This way the number of arguments of the wave functions is reduced and more importantly, the higher dimensional wave functions can be factorized into products of sine functions and the one dimensional wave functions, which are much easier to deal with than their higher dimensional counterparts. One distinct property of the 1 dimensional wave functions is that they are invariant under cyclic exchanges of odd numbers of particles, namely for $N$ odd, \begin{equation}
C_{+1}^x\Psi_{1D}(1,\cdots,N)=\Psi_{1D}(1,\cdots,N),
\end{equation}
where $C_{+1}^x$ represents the action to move every particle by one site in the $+x$ direction, with the last particle moved to the position of the first one, that is $1\to2,2\to3,\cdots,N\to1$.

Consider for example the non-degenerate ground state of 5 particles in 2 dimensions. For this state, it becomes straightforward to show that each group of the 3 near neighbors living in the real space square lattice are connected by products of four triplet exchanges, which are all performed along the 1 dimensional lines. Proven this, one can proceed as in the high temperature limit, by adding more particles to the lattice. And these newly added particles can be shown to be connected to the original particles' cluster by the similar method used for 5 particles. The only difference is that now one needs to consider the whole line of particles, on which the new particle is added, and thus a sequence of four cyclic exchanges, instead of the special triplet exchanges are required.
Since for non-degenerate ground states, there are odd number of particles on each line, cyclic exchanges will not produce extra minus signs, thus leading to the same result as triplet exchanges. This completes the proof for 2 dimensions, and the high dimensional cases are essentially the same.

However,  winding is a topological property that should be independent of representation. In the long time $\beta \rightarrow \infty$ limit the path integral
contains the same information as the ground state wave function, and for the Fermi-gas we can actually easily determine the winding properties inside one
of the nodal cells using the random permutation theory. This demonstrates that at zero temperature the Fermi-gas is indeed precisely equivalent to the Bose gas, within
the nodal cell.

\subsection{There is only room for winding at the bottom.}

The conclusion of the previous subsection is that the Ceperley wordlines can in principle  become infinitely long because the topology of
the reach allows them to become macroscopic. Does this mean that the zero-temperature order parameter of the Fermi-liquid is just
an algebraic  bose condensate of Ceperley wordlines characterized by a domination of the partition sum by macroscopic cycles? The
two nodal cell topological property is a necessary but  insufficient condition for this to be true. However, there are more reasons to believe
that the Fermi-liquid has to be of this kind. 

As we discussed at length in section II, the zero- and finite temperature Fermi-liquid are separated by a phase transition and it appears  that
only the winding sector of the Ceperley path integral can be responsible for this transition.  The argument is simple and general. With regard
to ordering dynamics the real space Ceperley path integral is governed  by
Boltzmannian principle and let us find out what 'substance' is available  to form an order parameter. The nodal surface in isolation cannot be responsible, since
it is an immaterial object that just governs the behavior of the 'Ceperley particles' . Henceforth, whatever its (singular) properties, 
these have to be reflected in the behavior of the matter.  In principle one can imagine subtle topological changes occurring in
the nodal surface but in the previous subsection we found this not to be the case  in the  Fermi-gas.
Henceforth, searching for the thermodynamic singularity we should keep our eyes on the worldlines and these should be subjected 
to the generalities associated with bosonic matter. One source of thermodynamic
singularity is that the system of bosons breaks the translational- and/or rotational symmetry of space, forming a crystal  or some liquid crystal. 
Although the one dimensional Fermi-gas is such a crystal in disguise, it is impossible to hide a (partial) crystallization in higher
dimensions: the higher dimensional Fermi-liquid is undoubtedly a true liquid. The worldlines have to be delocalized, but  
dealing with indistinguishable particles, being bosons or the 'even permuting' Ceperley particles,
one has to account for an extra set of degrees of freedom: the reconnections at the temporal boundary. From a statistical physics perspective, 
Bose condensation  appears as an order out of disorder phenomenon. Lowering temperature has the net effect of increasing the 'configurational
entropy' associated with all possible ways of reconnecting worldlines, or either the appearances of cycles characterized by different windings. 
Worldlines get longer and thereby the length over which they can meander increases, and this in turn increases effectively the fugacity of long cycles. 
The more cycles can contribute, the larger the 'configurational entropy' associated with the cycles and this gain in space time 'configurational entropy' 
(physically the  decrease of quantum zero point energy) causes eventually a flat distribution of the winding configurations, and in the Bose system this sets
in at a sudden phase transition. Since all particles 'are part of the same wordline' the Bose condensate is macroscopically coherent.
We learned that the reach allows  the Ceperley particles to form infinite windings. We learn from the Bose condensate that at zero temperature only
crystallization can prohibit the 'reconnection entropy' to take over, because the thermal de Broglie wavelength diverges. Henceforth, there does not
seem to be any feature of the reach that can prohibit this to happen as well  to the Ceperley worldlines at zero temperature. 

There is a quite direct argument to support this view which was put forward by Ceperley some time ago\cite{Cepstatphys,Cepx}. As we already emphasized a number of
times, on the canonical side the Fermi-liquid order manifests itself through the jump in the momentum distribution. Let us now turn to the
zero temperature  single particle density matrix,
\begin{eqnarray}
n ({\bf r}) & = &  \int d {\bf R} \rho ({\bf r}_1, {\bf r_2}, \cdots {\bf r}_N; {\bf r}_1 + {\bf r}, {\bf r_2}, \cdots {\bf r}_N; \infty)\nonumber\\
& = & \int \textrm{d} {\bf k} e^{i {\bf k} \cdot {\bf r}} n_{\bf k}.
%\label{partdens}
\end{eqnarray}

In the boson condensate $n_B ( {\bf r} ) \rightarrow constant$ revealing the off-diagonal long-range order which is equivalent to the domination of infinite
cycles. In the Femi-liquid on the other hand, 

\begin{equation}
n_F ( {\bf  r} )  \simeq  \frac{1}{(k_F r)^{d/2}} J_{d/2} (k_F r).
\label{partdensF}
\end{equation}

The oscillations governed by the Bessel function $J_{d/2} (k_F r)$  can be easily traced back to the size of the nodal pocket as discussed in a moment.
However, the envelope function $ (k_F r)^{-d/2}$ just behaves like the one particle density matrix of a Bose condensate showing off-diagonal 
long range order, like in the interacting Bose system in 1+1D at zero temperature.  Relating this to the real space Ceperley path integral, this signals the presence of 
infinite cycles formed from Ceperley world lines.       

Believing the arguments in the above, indicating that the Fermi liquid 'order' in the real space language is associated with the statistics of the windings,
we are still facing the problem why this behaves as a system of soft core, interacting bosons in 1+1D:  as we discussed in section II,
we learn from the canonical description that the 'reconnection entropy' of the Fermi-liquid in arbitrary dimensions at finite
temperature is counted in the same way as that of a soft-core (i.e. winding)  Bose system in 1+1 dimensions. This is surely not due to some remnant of
the 'supersymmetric' quantum statistics coming from the even windings because we learned in Section IV that this is quite similar to normal Bose statistics.
We suspect that this "holography" has its origin in  a non-topological, microscopic feature of the reach. 

It seems impossible to explain such a 'dimensional reduction' In terms of local particle-particle interactions and in one or the other way the reach apparently
acts  likely as a many particle interaction. An indication that this mysterious 'interaction' on microscopic time and length scales follows from the thermodynamics
of the Sommerfeld gas. We observe that temperature enters the  Eq.'s (\ref{freeensom})-(\ref{chpotsom})  always in the ratio $T/E_F$. 
Temperature has  the role of limiting the length of the imaginary time axis and this information enters the free energy via some effective return probabilities.
These in turn relate in a simple way with the Fermi-energy because otherwise the free energy would not be a simple algebraic function of $T/ E_F$, and the
meaning of the Fermi-energy is quite clear. In analogy with the one dimensional case,  the Fermi-energy is
the quantum kinetic energy associated with the fact that the free volume in which a Ceperley particle moves is restricted by the reach to be of order of the
inter-particle distance.  The reach is for a given ${\bf R}_0$ and $\tau$  just a $(Nd-1)$-dimensional manifold embedded in $Nd$ dimensional
space. It can be visualized by constructing d dimensional cuts through configuration space obtained by fixing $N-1$ particle coordinates and moving one
particle around, tracking the sign changes of the density matrix, as in Fig. 2.  For the Fermi-gas this is obviously a smooth, non-fractal manifold. We also know that the
Pauli hypersurface lies on the nodal surface. Given the typical inter-particle distance $r_s$, one expects that every particle moves around in a free volume with a 
typical linear dimension $\simeq r_s$, enclosed by the impenetrable nodal surface. This is the well known 'nodal pocket'; following Ceperley\cite{Cepstatphys,Cepx}, one can isolate
it  by analyzing the sign changes of the one particle density matrix.  But one might as well consider the meaning of the Fermi-energy.  
$E_F$ sets the ultraviolet scale of the higher dimensional Fermi gas in much the same way as it does in 1+1D. There we saw
that it corresponds to the time where the constraints become noticeable in the form of a confinement effect. The nodal pockets in higher dimensions have the same confining
effect as the 'Pauli pockets' in one dimension: the nodal surface 'hangs' over the Pauli surface forming 'cages'  in which the worldlines of individual particles are
locked up (see Fig 3). The derivation Eq.'s (\ref{msqfluc1D}-\ref{colener1D}) generalizes straightforwardly to any dimension after adjusting volume factors. The inter-particle
density $r_s = (V / N)^{1/d} = n^{1/d}$ where $n$ is the density. It follows for the mean square displacement $l_c^2 (\tau)$, the collision time $\tau_c$ and the
characteristic energy scale,

\begin{subequations}
\begin{eqnarray}
l_c^2 (\tau)  & =  & \langle ( \phi_i (\tau) - \phi_i (0) )^2 \rangle =  d  \frac {\hbar} {M} \tau, \\
\tau_c  & \simeq  & \frac{1}{d} \frac{M}{\hbar} n^{-2/d}, \\
E_F & = & \frac{\hbar}{\tau_c} = d \frac{\hbar^2} {2m} n^{2/d},
\label{higherdEF}
\end{eqnarray}
\end{subequations}
and one recognizes immediately that the characteristic energy scale coincides with the Fermi-energy. 

Let us now return to the question of what is responsible for the holography. The Fermi-energy emerges at microscopic times, and has a 
simple interpretation in terms of the microscopic geometry of the reach (the nodal pocket). Combining this observation with the fact that
the free energy  of the Fermi-gas at finite temperature is a simple algebraic function of the ratio $T / E_F$ suggests that the winding statistics
responsible for the thermodynamics is in some quite direct way influenced by the nodal pocket property of the reach. It is intriguing that the 
dimensionality of the nodal pocket is the same as the dimensionality of space, suggesting that it has in principle the capacity to render the
winding statistics to become independent of dimensionality, but we have not managed yet to find out how this works in detail.  The reader
is cordially invited to give this fascinating problem a deep thought.

\renewcommand{\theequation}{A-\arabic{equation}}
% redefine the command that creates the equation no.
\setcounter{equation}{0}  % reset counter 
\section*{APPENDIX}  % use *-form to suppress numbering

\subsection*{Windings vs. Loops}
\label{WvsC}

%We are interested here in studying free systems with a bose-fermi degeneracy that leads to supersymmetry.
The partition function in the canonical ensemble for such a system, given in terms of the density matrix as 
in \eqref{partition-function-1} and \eqref{pathstat}, with $V=0$, factorizes into 
a product of Gaussian convolution integrals\footnote{The details of this calculation can be found in
chapter 7 of \cite{Kleinert}.}.
The convolutions then extend the path lengths in the imaginary time direction of paths where
the end-points are permuted among each other, as seen in figure 1.
In the extended zone scheme such paths satisfy periodic boundary conditions, regardless of whether
the particles are bosons or fermions, and have path lengths which are integer multiples of the length of
the temporal direction (inverse temperature), $w \hbar \beta$, where the integer is interpreted as
a winding number. The bose/fermi nature of the particles is still present in an overall permutation sign
in summing over all permutations which generate the windings.
The contribution of each path of winding number $w$ to the partition function of $N$ particles is
the same as that of a single particle (which can only have winding number one) living at $w$ times
the inverse temperature
\begin{equation}
  \Delta Z_0^{(N),w} \: = \: Z_0 ( w \beta ) \ .
\end{equation}

We will now show that the partition function of a free system of $N$ particles at inverse
temperature $\beta$ can be written in terms of a sum over windings subject to a certain constraint,
with the partition function of a single particle appearing in the product.
There are altogether $N!$ permutations of the the $N$ particles. When decomposing these into cycles
we need to keep track of the number of cycles of length $w$ in a given permutation.
Let us denote, for a given permutation, the number of cycles in that permutation of length $w$, by $C_w$.
Then with each permutation is associated a series of numbers $C_w$, with $w=1,...,N$.
Then, a sum over all permutations can be rewritten as a sum over all integers assigned to
the various $C_w$, subject of course to an overall constraint, this constraint being that the total
length of all cycles taken together must be $N$. We write this constraint as
\begin{equation}
  \sum_{w=1}^N \: w \: C_w \: = \: N
\end{equation}
since the length of cycles in the class $C_w$ is $w$.
Among the $N!$ permutations, the number of which have a given set of values of $C_w$ has been given
already in \eqref{comb-factors}, which we reproduce here
\begin{equation}
  M(C_1,...,C_N) \: = \:
  \frac{N!}{\prod_{w=1}^N C_w ! \ w^{C_w}}
  \ .
\end{equation}
If we sum this quantity over all configurations of $C_w's$ such that
$\sum_{w=1}^N = L$, for any $1 \le L \le N$, at fixed $N$,
\begin{equation}
\label{cycle-1}
  \sum_{
    \begin{smallmatrix}
      \{C_1 , \cdots C_N \}  \\
      \tiny{\sum_{w=1}^N C_w = L}    
    \end{smallmatrix}
   }
  M(C_1,...,C_N) \: = \:
  S_1(N,L)
\end{equation}
where $S_1(N,L)$ is the Stirling number (unsigned) of the first kind, and counts the number of
permutations among all $N!$ permutations that consist of $L$ loops, for any number of loops
$1 \le L \le N$. Since there are $N!$ permutations,
summing this quantity over all allowed loop numbers must reproduce
\begin{equation}
\label{cycle-2}
  \sum_{L=1}^N \: S_1(N,L) \: = \: N!
  \ .
\end{equation}
Naturally, summing over all cycles of various lengths, we have the total number
of cycles
\begin{equation}
\label{cycle-3}
  \sum_{w=1}^N \: C_w \: = \: \text{Total number of cycles.}
\end{equation}

In writing the partition function in the canonical ensemble in terms of cycles we encounter
a sum of products of the form
\begin{equation}
  \sum_{\mathcal{P}} (\pm 1)^\mathcal{P}
  \prod_{
    \begin{smallmatrix}
      w=1
%  \\
%      \sum_w w C_w = N    
    \end{smallmatrix}
   }^N
  \ ,
\end{equation}
with $\mathcal{P}$ denoting a permutation.
Note here that the first sum over all permutations fixes the values of the $C_w$,
subject to the constraint on total length
$\sum_w w C_w = N$.
This sum of products can be rewritten as a sum over all values of the various $C_w$, again subject to the
constraint on total length
\begin{equation}
  \sum_{
    \begin{smallmatrix}
      \{ C_1 \cdots C_N \}  \\
      \sum_w w C_w = N    
    \end{smallmatrix}
   }
   \epsilon_w(C_1,\cdots,C_N)
%   (\pm 1)^{\sum_w (1+w) C_w}
  \prod_{w=1}^N
  \ ,
\end{equation}
where $\epsilon_w(C_1,\cdots,C_N) = (\pm 1)^{\sum_w (1+w) C_w}$ is the parity of a given permutation.
With these we can write the canonical partition function as \cite{Kleinert}
\begin{equation}
\label{canonical-pf-0}
  Z_0^{(N)} (\beta) \: = \:
  \frac{1}{N!} \:
  \sum_{
    \begin{smallmatrix}
      \{ C_1 \cdots C_N  \} \\
      \sum_w w C_w = N    
    \end{smallmatrix}
   }
  M
% (C_1,\cdots,C_N) \:
  \epsilon_w
%(C_1,\cdots,C_N) \:
  \prod_{w=1}^N
  \left[ Z_0(w \beta) \right]^{C_w}
  \ ,
\end{equation}
(dropping for clarity the $C_1,\cdots,C_N$ on $M$ and $\epsilon_w$).
After some starightforward reorganization this becomes
\begin{equation}
\label{canonical-pf}
  Z_0^{(N)} (\beta) \: = \:
  \sum_{
    \begin{smallmatrix}
      \{ C_1 \cdots C_N  \} \\
      \sum_w w C_w = N    
    \end{smallmatrix}
   }
  \prod_{w=1}^N
  \frac{1}{C_w!}
  \left[
    (\pm 1)^{w-1} \: \frac{Z_0 (w \beta)}{w}
  \right]^{C_w}
  \ .
\end{equation}
One may readily check that the partition function above
satisfies the recursion relation
\begin{equation}
\label{recursion-relation-canonical}
  Z_0^{(N)}(\beta) \: = \:
  \frac{1}{N} \: \sum_{n=1}^N \: (\pm1)^{n-1} \:
  Z_0^{(1)} (n \beta) Z_0^{(N-n)}(\beta) \ ,
\end{equation}
subject to the boundary condition $Z_0^{(0)}=1$, allowing one to generate the $N$-particle
partition functions from the bottom up.

In writing the partition function for an $N$-particle system as it appears in
\eqref{canonical-pf-0} and \eqref{canonical-pf}, we have taken the
viewpoint that we sum over all windings. Yet, it is evident in equations
\eqref{cycle-1} to \eqref{cycle-3} that we might also choose to view the same
partition function as a sum over all possible numbers of loops.
In other words, we may identify the sums as follows
\begin{equation}
  \sum_{
    \begin{smallmatrix}
      \{ C_1 \cdots C_N  \} \\
      \sum_w w C_w = N    
    \end{smallmatrix}
   }
  \: = \:
  \sum_{R=1}^N \:
  \sum_{
    \begin{smallmatrix}
      \{ C_1 \cdots C_N  \} \\
      \sum_w C_w = R \\
      \sum_w w C_w = N
    \end{smallmatrix}
   }
\end{equation}
The second sum on the right hand side is subject to two constraints. But we will see shortly that when
we construct the grand canonical partition function, the constraint
$\sum_w w C_w = N$ will be removed, since we sum over all canonical partition functions for the various particle numbers, leaving only the one enforcing the number of loops $\sum_w C_w = R$.

Define the partition function for $N$-particles and $R$-loops as
\begin{equation}
\label{define-pf-nr}
  \tilde{Z}_{0,R}^{(N)} (\beta) =
  \sum_{
    \begin{smallmatrix}
      \{ C_1 \cdots C_N  \} \\
      \sum_w C_w = R \\
      \sum_w w C_w = N
    \end{smallmatrix}
   }
  M
% (C_1,\cdots,C_N)
  \prod_{w=1}^N
  \left[ Z_0^{(1)}(w \beta) \right]^{C_w}
  \ .
\end{equation}
Note that no sign appears here, and so this gives the same result whether we are studying
bosons or fermions.
We can now perform the winding to loop regrouping on \eqref{canonical-pf-0}, to see that
\begin{eqnarray}
\label{canonical-pf-cycles}
  Z_0^{(N)}(\beta) \: = \:
  \frac{(\pm1)^N}{N!} \:
  \sum_{R=1}^N \: (\pm1)^R \:
  \tilde{Z}_{0,R}^{(N)} (\beta)
  \ .  
\end{eqnarray}
We note that by going to this view, we have greatly reduced the number of alternating
signs in the sums that we have to deal with, from $N!$ signs in \eqref{canonical-pf-0}
to $N$ sings in \eqref{canonical-pf-cycles}. For those interested in numerical
simulations this from of the partition function might offer a substantial
improvement in the rate of convergence.

The analogue of the recursion relation \eqref{recursion-relation-canonical} is then
cycle form is
\begin{eqnarray}
  \tilde{Z}_{0,R}^{(N)} (\beta) & = & \frac 1N \sum_{n=1}^{N-1} Z_0(n\beta) \tilde{Z}_{0,R-1}^{(N-n)} (\beta)\quad (R\ge 2),\nonumber\\
   \tilde{Z}_{0,1}^{(N)} (\beta) & = & \frac 1N Z_0(N\beta).
\end{eqnarray}

We are now going to use the canonical partition function to construct the grand canonical one, where
the particle number is no longer a fixed, but a fluctuating quantity.
The grand canonical partition function is the sum over all canonical partition functions of different
particle numbers, each weighted by a Boltzmann for $N$ particles, and at chemical potential $\mu$,
\begin{equation}
  Z_{G,0} \: \equiv \:
  \sum_{N=0}^\infty \: Z_0^{(N)} (\beta) \: e^{\beta \mu N}
  \ ,
\end{equation}
with the free partition function satisfying
\begin{equation}
  Z_{G,0} ( w \beta ) \: = \: \frac{Z_{G,0}(\beta)}{w^{\frac{d}{2}}} \ .
\end{equation}

Using the form of the canonical partition function in \eqref{canonical-pf}, we have then for the
grand canonical partition function, after a rearrangement, the following
\begin{equation}
\label{grand-pf}
  Z_{G,0} (\beta) \: = \:
  \prod_{w=1}^\infty
  \sum_{C_w = 0}^\infty
  \frac{1}{C_w !}
  \left[
    (\pm 1)^{w-1} \: \frac{Z_0 (w \beta) e^{w \beta \mu}}{w}
  \right]^{C_w}
  \ ,
\end{equation}
where because we sum over all partical numbers, the constraint in \eqref{canonical-pf} has been removed
(as promised), i.e.
\begin{equation}
  \sum_{N=0}^\infty \: 
  \sum_{
    \begin{smallmatrix}
      \{ C_1 \cdots C_N  \} \\
      \sum_w w C_w = N    
    \end{smallmatrix}
   }
   \: = \:
  \prod_{w=1}^\infty
  \sum_{C_w = 0}^\infty
  \ .   
\end{equation}
In terms of cycles, the removal of the constraint means we have
\begin{equation}
  \sum_{N=0}^\infty \: 
  \sum_{
    \begin{smallmatrix}
      \{ C_1 \cdots C_N  \} \\
      \sum_w w C_w = N    
    \end{smallmatrix}
   }
   \: = \:
  \sum_{R=1}^\infty \:
  \sum_{
    \begin{smallmatrix}
      \{ C_1 \cdots C_N  \} \\
      \sum_w C_w = R
    \end{smallmatrix}
   }
\end{equation}

\section {Acknowledgments}

We acknowledge insightful discussions with D.I. Ceperley.
This research was supported by the "Nederlandse organisatie voor Wetenschappelijk Onderzoek"  (NWO) and 
by the "Stichting voor Fundamenteel Onderzoek der Materie" (FOM).

\bibliographystyle{apsrev}

\end {document}